\documentclass[journal]{IEEEtran}

\usepackage{cite}
\usepackage{color}
\usepackage[pdftex]{graphicx}
\graphicspath{{Figs/}{../pdf/}{../jpg/}}
\DeclareGraphicsExtensions{.pdf,.jpg,.png}
\usepackage{lipsum}
\usepackage{amsmath}
\usepackage{amssymb}
\usepackage{mathrsfs}
\usepackage{algorithm}
\usepackage{array}
\usepackage{fixltx2e}
\usepackage{stfloats}
\usepackage{url}
\usepackage{booktabs}
\usepackage{enumerate}
\usepackage{setspace}

\usepackage{amsmath}
\usepackage{amsthm}
\newtheorem{theorem}{Theorem}[section]

\newtheorem{proposition}[theorem]{Proposition}

\newtheorem{remark}[]{Remark}

\definecolor{CBLUE}{RGB}{0,114,189}
\definecolor{CRED}{RGB}{217,83,25}
\definecolor{CYELLOW}{RGB}{237,177,32}
\definecolor{CPURPLE}{RGB}{126,47,142}

\usepackage[normalem]{ulem}
\newcommand\rout{\bgroup\markoverwith{\textcolor{red}{/}}\ULon} 

\usepackage[prependcaption,colorinlistoftodos]{todonotes}

\begin{document}

\title{$\mathcal{H}_{\infty}$-Control of Grid-Connected Converters: Design, Objectives and Decentralized Stability Certificates}

\author{Linbin Huang, Huanhai Xin, and Florian D{\"o}rfler
\thanks{L. Huang is with the College of Electrical Engineering at Zhejiang University, Hangzhou, China, and the Department of Information Technology and Electrical Engineering at ETH Zurich, Switzerland. (Email: \text{huanglb@zju.edu.cn})}
\thanks{H. Xin is with the College of Electrical Engineering at Zhejiang University, Hangzhou, China. (Email: \text{xinhh@zju.edu.cn})}
\thanks{F. D{\"o}rfler is with the Department of Information Technology and Electrical Engineering at ETH Zurich, Switzerland. (Email: \text{dorfler@ethz.ch})}
\thanks{This research was supported in part by the 2020 Science and Technology Project of State Grid Power Company Limited (``Power Grid Strength Evaluation and Optimization to Accommodate High-penetration Renewables'') and in part by ETH Zurich Funds.}}


\maketitle


\begin{abstract}

The modern power system features high penetration of power converters due to the development of renewables, HVDC, etc. Currently, the controller design and parameter tuning of power converters heavily rely on rich engineering experience and extrapolation from a single converter system, which may lead to inferior performance or even instabilities under variable grid conditions. In this paper, we propose an $\mathcal{H}_{\infty}$-control design framework to provide a systematic way for the robust and optimal control design of power converters. We discuss how to choose weighting functions to achieve anticipated and robust performance with regards to multiple control objectives. Further, we show that by a proper choice of the weighting functions, the converter can be conveniently specified as grid-forming or grid-following in terms of small-signal dynamics. Moreover, this paper first proposes a decentralized stability criterion based on the small gain theorem, which enables us to guarantee the global small-signal stability of a multi-converter system through local control design of the power converters. We provide high-fidelity nonlinear simulations and hardware-in-the-loop (HIL) real-time simulations to illustrate the effectiveness of our method.
\end{abstract}


\section{Introduction}

The penetration rate of power-electronic converters in the modern power system is ever-increasing mainly because of the rapid development of renewables, energy storage systems and high-voltage DC transmission (HVDC) systems. The high controllability of power converters is making the power system more flexible, which allows auxiliary services such as frequency support, voltage support and oscillation damping to be provided for the power grid \cite{rocabert2012control,FM-FD-GH-DH-GV:18}.

Commonly, power converters facilitate multiple control loops to achieve different control objectives. For example, in a typical grid-following converter, a phase-locked loop (PLL) is used for grid synchronization; a current control loop is used for current control and fast current limitation; an active power control loop is used for active power tracking; and a voltage control loop is used to regulate the terminal voltage \cite{suul2016impedance}. The corresponding control structure, i.e., how these loops are interconnected and tuned, is currently based on analysis of a single converter together with rich engineering experience and physical intuition. Compared with conventional synchronous generators, power converters generally present much more complex dynamics in a wide frequency range due to these multiple loops, thereby posing great challenges to the analysis, operation, and control of future low-inertia power grids \cite{FM-FD-GH-DH-GV:18}.

It has been revealed that the control loops inside a converter are strongly coupled with each other, deteriorate the performance and even lead to instabilities under variable grid conditions \cite{suul2016impedance,wang2018unified,huang2019grid,zhang2017sequence}. The coupling among different loops complicates the parameter tuning of converters as proper parameters can hardly be directly obtained from a tractable model. The most common method to deal with this issue is to model the whole converter system, analyze how certain ranges of the parameters affect the stability of the converter (through eigenvalue loci, Nyquist diagrams, etc.), and then pick some acceptable parameter sets to be tested in a real system \cite{wen2016analysis,pogaku2007modeling,suul2009design}. However, this method may not lead to the optimal parameter set in terms of damping ratio or stability margin because eigenvalue loci or Nyquist diagrams can hardly deal with multiple parameters (e.g., the converter may have eight parameters to tune if it has four loops). It is even more challenging to design optimal and stabilizing controller with regards to multi-converter systems, as the converters can be strongly coupled and lead to complex interacting dynamics.

One convenient way to achieve optimal performance is to use $\mathcal{H}_2/\mathcal{H}_{\infty}$-optimal control synthesis. In \cite{steenis2012robust} and \cite{steenis2013h}, $\mathcal{H}_{\infty}$ and gain scheduled $\mathcal{H}_{\infty}$ controllers were used in microgrids to perform robust control. However, only a simplified model of the converter (without current control loop) is considered, which may result in inferior performance for converters that have a current control loop.
An optimal voltage control problem for islanded power converters was posed in \cite{johnson2018optimal} and solved using $\mathcal{H}_{\infty}$ synthesis. In \cite{yang2010robust}, $\mathcal{H}_{\infty}$ synthesis was used to design the converter's output current controller and improve the robustness against variable grid impedance. In terms of harmonic suppression, $\mathcal{H}_{\infty}$ and repetitive control techniques can be used for designing output voltage controllers to reject harmonic disturbances from nonlinear loads or the public grid, as presented in \cite{weiss2004h}.

The above methods will result in a dynamic controller whose order is the same as that of the system, which further complicates the system dynamics. It is also possible to obtain {\em fixed-structure fixed-order} optimal controllers by $\mathcal{H}_2/\mathcal{H}_{\infty}$ synthesis \cite{gencc2003state}. For example, a robust frequency control was obtained in \cite{misyris2018robust} through $\mathcal{H}_{\infty}$ loop shaping design of a static control gain matrix to improve the grid frequency dynamics. A fixed-structure current controller based on $\mathcal{H}_{\infty}$ synthesis was developed in \cite{kammer2018convex} to guarantee robust stability and performance for power converters.
In \cite{sanchez2019optimal}, an $\mathcal{H}_2$-optimal tuning method was applied to optimize the PI gains in HVDC systems. In \cite{poolla2019placement}, $\mathcal{H}_{2}$-optimal design is applied to allocate the virtual inertia of grid-forming and grid-following converters in low-inertia systems. It is noteworthy that grid-forming converters and grid-following converters, which present distinct dynamic behaviors, have been widely integrated in modern power grids. It was shown in \cite{Marie2019Quantitative} that grid-forming and grid-following converters can be distinguished by their small-signal responses to grid frequency disturbances in the frequency domain. Based on this finding, we will analyze how to specify a converter as grid-forming or grid-following with regards to small-signal dynamics  for the purpose of guiding the $\mathcal{H}_{\infty}$-control design. Note that in this paper, akin to \cite{Marie2019Quantitative}, we consider PLL-based converter as one prototypical grid-following type and droop-controlled converters as one prototypical grid-forming type.

Both $\mathcal{H}_2$ and $\mathcal{H}_{\infty}$ optimal controllers are capable of stabilizing the system. However, the existing $\mathcal{H}_2/\mathcal{H}_{\infty}$-optimal design for power converters generally considers a single-converter system, which can only ensure the system stability when the converter is connected to an infinite bus, but provides no guarantee on the stability of a multi-converter system.
Moreover, the control objective of stable grid synchronization under variable grid conditions was not included in the $\mathcal{H}_2/\mathcal{H}_{\infty}$-optimal design in the existing literature, which may result in inferior synchronization performance and poor robustness.

In this paper, we propose a fixed-structure $\mathcal{H}_{\infty}$-control design framework to perform optimal, robust, and multivariable control for grid-connected power converters. Our $\mathcal{H}_{\infty}$-control design considers multiple control objectives (grid synchronization, active power and voltage regulations) simultaneously. We elaborate on how to achieve the specified control objectives by choosing proper weighting functions. Moreover, we show that by choosing different weighting functions, the converter can be conveniently specified as grid-forming or grid-following in terms of small-signal dynamics. Furthermore, we propose a decentralized stability criterion based on the small gain theorem, which enables to ensure the global stability of a multi-converter system through local $\mathcal{H}_{\infty}$-control design of the converters. We show that the resulting $\mathcal{H}_{\infty}$-optimal controller has the anticipated dynamic performance, robustness against variable grid conditions, and it guarantees on the overall system stability. We illustrate our results by means of high-fidelity simulations and hardware-in-the-loop (HIL)  real-time simulations.

The rest of this paper is organized as follows: Section~II presents the $\mathcal{H}_{\infty}$-control design framework for the grid-connected power converters. Section~III discusses the control objectives and the corresponding weighting functions. Section~IV proposes the decentralized stability criterion for multi-device systems and shows how it can be incorporated in the $\mathcal{H}_{\infty}$-control design framework. Detailed simulation and HIL real-time simulation results are provided in Section~V. Section~VI concludes the paper.

\section{$\mathcal{H}_{\infty}$-Controller for Converters}
\label{section Hinf controller}

In this section we present the design setup of our fixed-structure $\mathcal{H}_{\infty}$-controller for power converters, and compare its structure with conventional control schemes.


\subsection{Converter system descriptions}

\begin{figure*}[!t]
	\centering
	\includegraphics[width=6.6in]{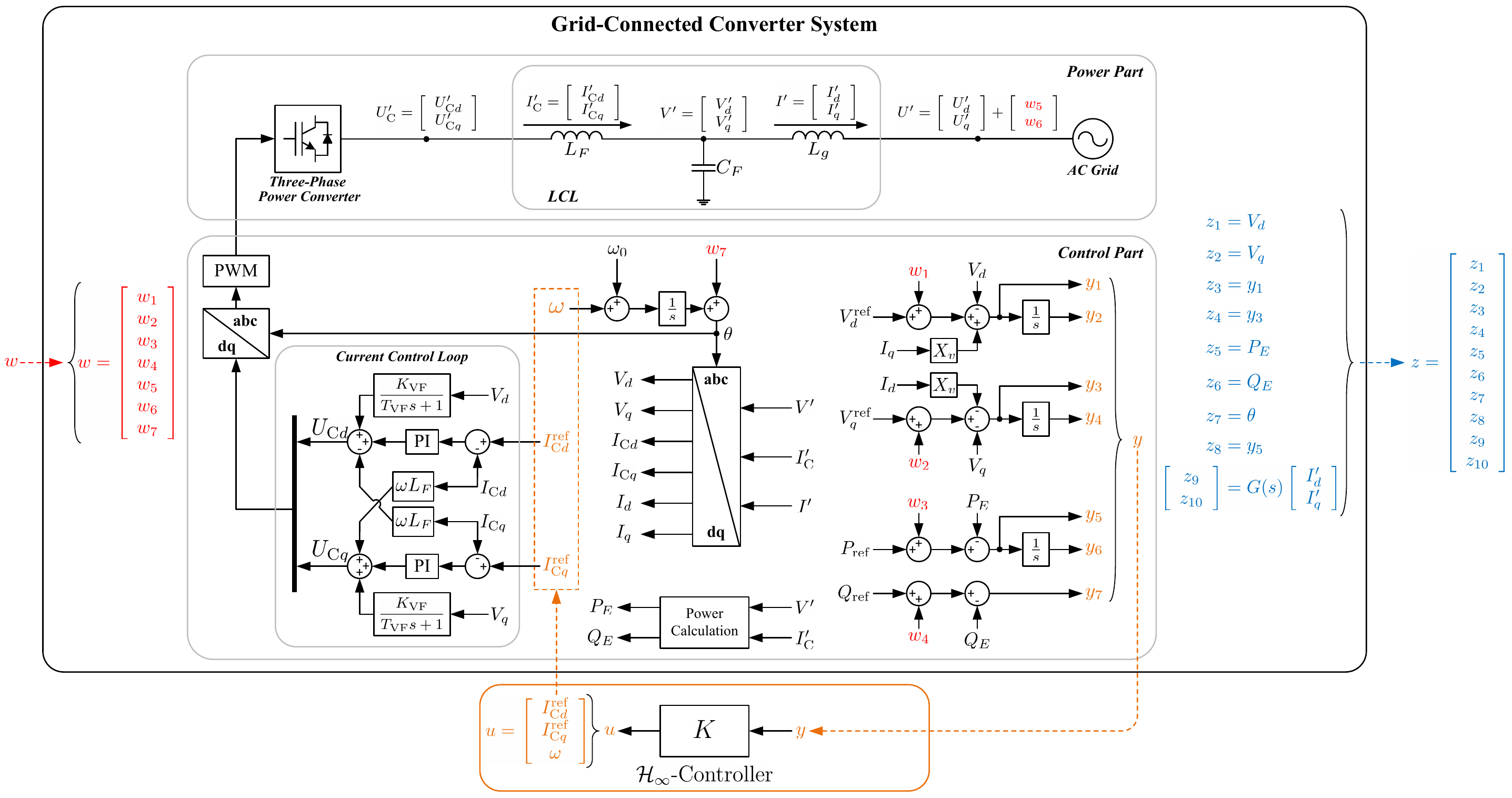}
	\vspace{-4mm}
	\caption{One-line diagram of a three-phase power converter that is connected to the power grid via an {\em LCL} ($L_F$, $C_F$ and $L_g$ are the {\em LCL} parameters).}
	\vspace{-2mm}
	\label{Fig_Control_diagram}
\end{figure*}

Fig.~\ref{Fig_Control_diagram} shows a three-phase power converter which is connected to a power grid via an {\em LCL} filter. In the control scheme of this converter, the three-phase voltage and current signals are represented by two-dimensional vectors in the synchronously rotating {\em dq} frame (through Park transformation). Usually, the rotating frequency of this {\em dq} frame is generated by a grid-synchronization unit, e.g., PLL, emulated swing equation or droop control. Here this frequency (denoted by $\omega + \omega_0$, and $\omega_0$ is the nominal value) is generated by the $\mathcal{H}_{\infty}$-controller, as depicted in Fig.~\ref{Fig_Control_diagram}.
A current control loop is used to make the converter-side current ($I_{{\rm C}d}$ and $I_{{\rm C}q}$) track their reference values ($I_{{\rm C}d}^{\rm ref}$ and $I_{{\rm C}q}^{\rm ref}$) and implement fast current limitation \cite{xin2016synchronous}.
In the global {\em dq} frame (with constant rotating frequency $\omega_0$), the converter-side voltage vector is denoted by $U'_{\rm C}$, which is determined by the PWM signals; the converter-side current vector is $I'_{\rm C}$; the grid-side current vector is $I'$; the capacitor voltage vector of the {\em LCL} filter is $V'$; and the grid-side voltage vector is $U'$, as labeled in Fig.~\ref{Fig_Control_diagram}.

Let $Y(s)$ be the converter's admittance matrix (linearized model), which is a $2 \times 2$ transfer function matrix reflecting how a perturbation from the terminal voltage affects the converter's current output, i.e., $I' = Y(s)U'$. It has been demonstrated that this admittance matrix dominates the stability of a converter system since it describes the input/output characteristics of the converter as seen from the point of the AC grid, and for the detailed derivation process we refer to \cite{wang2018unified}, \cite{wen2016analysis} or \cite{huang2019impacts}.

\vspace{-1.5mm}

\subsection{$\mathcal{H}_{\infty}$-control design setup}

The converter system in Fig.~\ref{Fig_Control_diagram} {(not including the $\mathcal{H}_{\infty}$-controller $K$)} can be modeled as
\begin{equation}
\left[\begin{array}{*{20}{c}}
z\\
y
\end{array}\right] = \left[\begin{array}{*{20}{c}}
P_{11}(s)&P_{12}(s)\\
P_{21}(s)&P_{22}(s)
\end{array}\right]\left[\begin{array}{*{20}{c}}
w\\
u
\end{array}\right]\,,
\label{eq:zy}
\end{equation}
where $u$ is the control input of the system, $y$ is the measured output for the $\mathcal{H}_{\infty}$-controller, $w$ and $z$ are the input/output signals chosen to quantify the performance of the system \cite{skogestad2007multivariable}. Fig.~\ref{Fig_Control_diagram} shows how the elements of $w$ enter the system as disturbances, and how $z$ is chosen from the system variables.
To quantify the power and voltage tracking performance, we choose $w_1$, $w_2$, $w_3$ and $w_4$ to be the disturbance inputs imposed on the power and voltage reference signals, as shown in Fig.1. The disturbances $w_5$ and $w_6$ are used to incorporate the admittance matrix in the $\mathcal{H}_{\infty}$-controller for decentralized stability certification, which will be elaborated upon in Section IV. The disturbance signal $w_7$ is used to specify the grid-synchronization performance of the system, as grid synchronization requires the angle output (i.e., $\theta$) to reject disturbances that enter the closed loop.

In this paper, $u$ is chosen as $u=\left[I_{{\rm C}d}^{\rm ref}\;\;I_{{\rm C}q}^{\rm ref}\;\;\omega\right]^\top$, and $y$ is the vector of voltage/power signals and their integrals
\begin{equation}
y = \left[ {\begin{array}{*{20}{c}}
	y_1\\
	y_2\\
	y_3\\
	y_4\\
	y_5\\
	y_6\\
	y_7
	\end{array}} \right] = \left[ {\begin{array}{*{20}{c}}
	V_d^{\rm ref} - V_d + I_qX_v + w_1\\
	\frac{1}{s}(V_d^{\rm ref} - V_d + I_qX_v + w_1)\\
	V_q^{\rm ref} - V_q - I_dX_v + w_2\\
	\frac{1}{s}(V_q^{\rm ref} - V_q - I_dX_v + w_2)\\
	P_{\rm ref} - P_E + w_3\\
	\frac{1}{s}(P_{\rm ref} - P_E + w_3)\\
	Q_{\rm ref} - Q_E + w_4\\
	\end{array}} \right]\,,
\label{eq:y}
\end{equation}
where $V_d^{\rm ref}$ and $V_q^{\rm ref}$ are {\em d}-axis and {\em q}-axis voltage references, $V_d$ and $V_q$ are the {\em d}-axis and {\em q}-axis voltage components, $P_{\rm ref}$ and $Q_{\rm ref}$ are active and reactive power references, $P_E$ and $Q_E$ are converter's active and reactive power, and $X_v$ is the virtual reactance to enhance the system flexibility. In addition to the virtual impedance loop \cite{rocabert2012control}, other auxiliary control loops (e.g., for damping oscillations \cite{rodriguez2019coupling}, improving power quality \cite{wang2015grid}, etc.) can also be conveniently included in the $\mathcal{H}_{\infty}$-control design framework by modifying the power/voltage references or the current control loop, which changes the model in \eqref{eq:zy} and possibly also the specification of the signals $(u,y,w,z)$.

\begin{remark}[Alternative outputs]
 We remark that the integrals in $y$ determine the system's steady state. To be specific, the integrals of voltage and active power signals regulate the voltage and active power to their references in steady state. The integral of reactive power is not included in $y$ since the steady-state value of reactive power is determined by the power flow when the voltage is regulated to the reference value.
Also, other expected steady states, e.g., voltage droop, can be conveniently configured by changing the integrals in $y$.
\label{Alternative_outputs}
\end{remark}

\vspace{-3mm}

The aforementioned voltage regulation is similar to some existing voltage control loops (e.g., the ac voltage control loop in droop-controlled converters \cite{rocabert2012control,huang2017transient}) in the sense that the $dq$-axis components of $V'$ in the controller's rotating coordinate are regulated to their reference values. The frequency of the controller's coordinate (i.e., $\omega$) is determined by the $\mathcal{H}_{\infty}$-controller to achieve grid synchronization, which will be elaborated upon below.

The performance output vector $z$ is chosen as
\begin{equation}
\begin{split}
z & = \left[z_1\;\;z_2\;\;z_3\;\;z_4\;\;z_5\;\;z_6\;\;z_7\;\;z_8\;\;z_9\;\;z_{10}\right]^\top\\
&= \left[V_d\;\;V_q\;\;y_1\;\;y_3\;\;P_E\;\;Q_E\;\;\theta\;\;y_5\;\;[I'_d\;\;I'_q]G(s)^\top\right]^\top\,,
\end{split}
\label{eq:z}
\end{equation}
which quantifies the tracking and disturbance-rejection performance of the voltage, power and angle signals. Note that $G(s)$ is a $2 \times 2$ transfer matrix chosen to enable a decentralized stability criterion, which will be elaborated in Section \ref{Section Decentralized Stability Certificates}.

Note that the disturbance signals $w_1 \sim w_7$ and the performance output vector $z$ are chosen for specifying the control objectives and quantifying the system performance in the $\mathcal{H}_{\infty}$ control design. In other words, they are chosen in order to derive a proper transfer function matrix in \eqref{eq:zy} which describes the response characteristics of the system. On this basis, we can use weighting functions to shape the characteristics of the system and achieve certain control objectives, which will be detailed in the next section.

Finally, we remark that the transfer function matrix in \eqref{eq:zy} can be conveniently obtained by first deriving the state-space model of the system with the  inputs/outputs specified in Fig.~\ref{Fig_Control_diagram} and then transforming this state-space model to the input/output transfer function matrix. Another way is to  derive the transfer function matrix by incorporating the input/output configuration (defined in Fig.~\ref{Fig_Control_diagram}) and then modifying the existing frequency-domain models (e.g., those in \cite{huang2019grid} and \cite{wen2016analysis}) of power converters (including the dynamics of {\em LCL}, current control loop, etc.).

The $\mathcal{H}_{\infty}$-controller can be formulated as
\begin{equation}
u=Ky
\label{eq:Static_K}\,,
\end{equation}
where $K \in \mathbb{R}^{3 \times 7}$ is a static parameter matrix.

\vspace{4mm}
\noindent {\em Two Noteworthy Cases}
\vspace{2mm}

The controller in \eqref{eq:Static_K} can be regarded as a generalized form of PLL-based controller (grid-following type) and frequency droop controller (grid-forming type). For example, we obtain a typical PLL-based controller \cite{huang2019grid} for the converter by setting
\begin{equation}
K \!=\!
{\footnotesize
	\left[ {\begin{array}{*{20}{c}}\!\!
		{0}&{0}&{0}&{0}&{K_{\rm PP}}&{K_{\rm PI}}&{0}\!\!\\\!\!
		{-K_{\rm VP}}&{-K_{\rm VI}}&{0}&{0}&{0}&{0}&{0}\!\!\\\!\!
		{0}&{0}&{-K_{\omega \rm P}}&{-K_{\omega \rm I}}&{0}&{0}&{0}\!\!
		\end{array}} \right]},
\label{eq:PLL_based_controller}
\end{equation}
where $K_{\rm PP}$ and $K_{\rm PI}$ are the PI gains of the active power control loop, $K_{\rm VP}$ and $K_{\rm VI}$ are the PI gains of the voltage control loop, $K_{\omega \rm P}$ and $K_{\omega \rm I}$ are the PI gains of the PLL ($V_q^{\rm ref}$ can be set as zero to achieve voltage orientation).

Moreover, a frequency droop controller \cite{simpson2013synchronization,xin2016synchronous} (with a cascaded voltage/current control structure) can be obtained by
\begin{equation}
K =
{\footnotesize
	\left[ {\begin{array}{*{20}{c}}
		{K_{\rm VP}}&{K_{\rm VI}}&{0}&{0}&{0}&{0}&{0}\\
		{0}&{0}&{K_{\rm VP}}&{K_{\rm VI}}&{0}&{0}&{0}\\
		{0}&{0}&{0}&{0}&{K_f}&{0}&{0}
		\end{array}} \right]}\,,
\label{eq:Frequency_droop_controller}
\end{equation}
where $K_f$ is the frequency droop coefficient.

\subsection{$\mathcal{H}_{\infty}$-control formulation}

In the following, we discuss how $K$ can be obtained optimally by solving an $\mathcal{H}_{\infty}$-optimization problem so as to make the converter have optimal and multivariable performance.

By combining \eqref{eq:zy} and \eqref{eq:Static_K} we obtain the closed-loop transfer function matrix of the system as
\begin{equation}
\begin{split}
z & = \left\{ P_{11}(s)+P_{12}(s)K \left[I-P_{22}(s)K\right]^{-1}P_{21}(s) \right\}w\\
& \buildrel \Delta \over = P(K)(s)w\,,
\end{split}
\end{equation}
which indicates that the design of $K$ will affect the closed-loop performance of the system.

The standard $\mathcal{H}_{\infty}$-optimal control problem is to find a stabilizing controller $K$ by solving
\begin{equation}
\mathop {\min }\limits_K \|W(s)P(K)(s)\|_{\infty} = \mathop {\min }\limits_K \mathop {\max }\limits_\omega \bar \sigma\left[W(j\omega)P(K)(j\omega)\right]\,,
\label{eq:standard_Hinf}
\end{equation}
where $\|\cdot\|_{\infty}$ denotes the $\mathcal{H}_{\infty}$ norm, $W(s)$ is the user-defined diagonal weighting transfer function matrix, and $\bar \sigma(\cdot)$ denotes the largest singular value.

However, in \eqref{eq:standard_Hinf}, the transfer functions from all the inputs to a particular output share the same weighting function, which is not a suitable setting for the multi-objective design in this paper. For example, sometimes we expect $z_1$, which is the {\em d}-axis voltage $V_d$, to track a disturbance in the reference $w_1$, and meanwhile rejects the other disturbances in $w$. However, these objectives cannot be achieved simultaneously when $z_1$ has the same weighting function for all the inputs in $w$. In addition, standard algorithms to solve \eqref{eq:standard_Hinf} result in a high-dimensional dynamic controller with the same order as that of the system \cite{doyle1989state}, which may complicate the system dynamics.

To deal with these problems, in this paper, we consider $K$ as a matrix of static gains to ensure the simplicity and thus implementability of the resulting controller. The corresponding $\mathcal{H}_{\infty}$-optimal control problem can be solved with the algorithm in \cite{apkarian2006nonsmooth}. Moreover, in order to achieve multi-objective design, we specify the shape of $P(K)(s)$ by solving
\begin{equation}
\begin{split}
&\mathop {\min }\limits_K \|\mathcal W(s) \circ P(K)(s)\|_{\infty}\\
=& \mathop {\min }\limits_K \mathop {\max }\limits_\omega \bar \sigma\left[\mathcal W(j\omega) \circ P(K)(j\omega)\right]\,,
\label{eq:static_K_entrywise_W}
\end{split}
\end{equation}
where $\circ$ denotes the entrywise product of matrices, and $\mathcal W(s)$ is the weighting transfer function matrix (not necessarily diagonal) which has the same dimension as $P(K)(s)$.
The entrywise weighting functions in \eqref{eq:static_K_entrywise_W} enable us to specify the shape of every entry of $P(K)(s)$ and thus provide more flexibility than \eqref{eq:standard_Hinf}.
We will discuss the design of the weighting transfer function matrix $\mathcal W(s)$ in next section.

\subsection{Grid-forming and grid-following dynamics}

\begin{figure}[!t]
	\centering
	\includegraphics[width=2.6in]{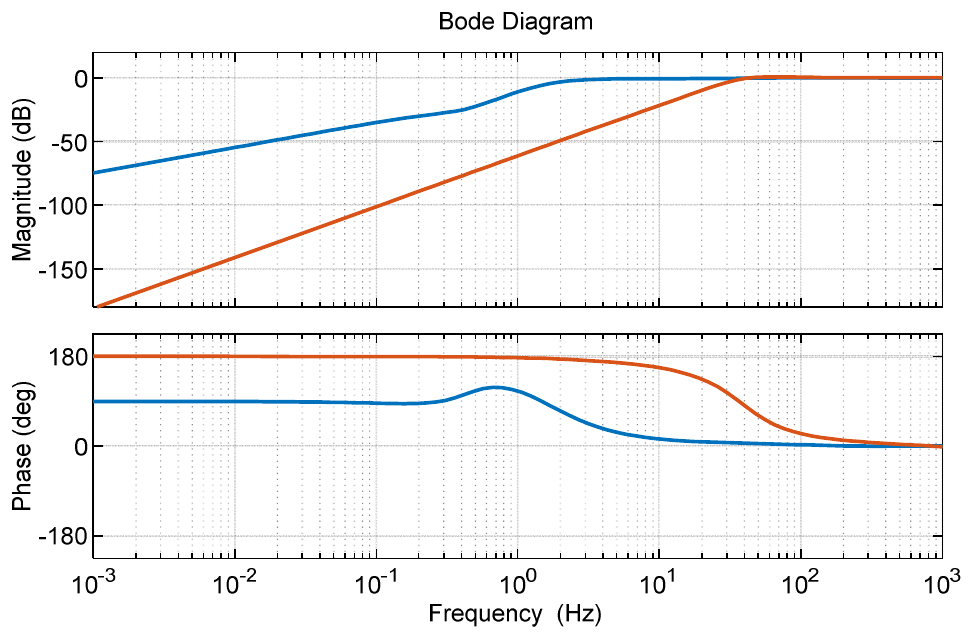}
	\vspace*{-2mm}
	\caption{Bode diagram of the sensitivity $P_{77}(K)(s)$. {\color{CBLUE}{\bf{-----}}} Droop controller by choosing $K$ as \eqref{eq:Frequency_droop_controller} and setting $K_{\rm VP} = 2$, $K_{\rm VI} = 10$ and $K_f = 4\pi$. {\color{CRED}{\bf{-----}}} PLL-based controller by choosing $K$ as \eqref{eq:PLL_based_controller} and setting $K_{\rm PP} = 0.5$, $K_{\rm PI} = 40$, $K_{\rm VP} = 0.5$, $K_{\rm VI} = 40$, $K_{\omega \rm P} = 171.8$ and $K_{\omega \rm I} = 14754.2$.}
	\vspace*{-2mm}
	\label{Fig_Angle_disturbance}
\end{figure}

The different settings for $K$ will inevitably result in different dynamic behaviors, e.g., making the converter behave like the grid-following or grid-forming type in terms of small-signal frequency/angle tracking dynamics. Fig.~\ref{Fig_Angle_disturbance} plots the closed-loop Bode diagrams of $P_{77}(K)(s)$ under droop control \eqref{eq:Frequency_droop_controller} and under PLL-based control \eqref{eq:PLL_based_controller}. Note that $P_{77}(K)(s)$ reflects how the controller rejects angle disturbances (from the grid) via the closed loop, which can be thought of as one single-input-single-output sensitivity function of the system \cite{skogestad2007multivariable}.

Observe that the PLL-based controller has a higher bandwidth than the droop controller, which indicates that the PLL-based controller has higher tracking speed but at the same time also higher sensitivity to grid disturbances. These observations are consistent with \cite{Marie2019Quantitative} analyzing the complementary sensitivity and showing that grid-following {converters generally have higher} control bandwidth to track the grid frequency.

\section{Control Objectives and Weighting Functions}
\label{subsection Design of weighting transfer function matrix}

When solving the $\mathcal{H}_{\infty}$-optimal control problem, the weighting transfer function matrix $\mathcal W(s)$ is used to specify the shape of $P(K)(s)$ and thus the performance objectives of the system. To be specific, the entry of $\mathcal W(s)$ in the $i {\rm th}$ row and $j {\rm th}$ column, i.e., $\mathcal W_{ij}(s)$, shapes how the $i{\rm th}$ output is affected by the $j{\rm th}$ input, i.e., the transfer function $P_{ij}(K)(s)$.


\subsection{Brief review on weighting functions for $\mathcal{H}_{\infty}$-control}
\label{subsec: weighting functions}

The weighting functions {$\mathcal W(s)$} in $\mathcal{H}_{\infty}$-control design in \eqref{eq:static_K_entrywise_W} can be in fact considered as the expected upper bounds for shaping the transfer function matrix of the system $P(K)(s)$. To illustrate this point, we explain how to choose a weighting function to specify reference tracking or disturbance rejection from a particular input to a particular output.

\begin{figure}[!t]
	\centering
	\includegraphics[width=2.6in]{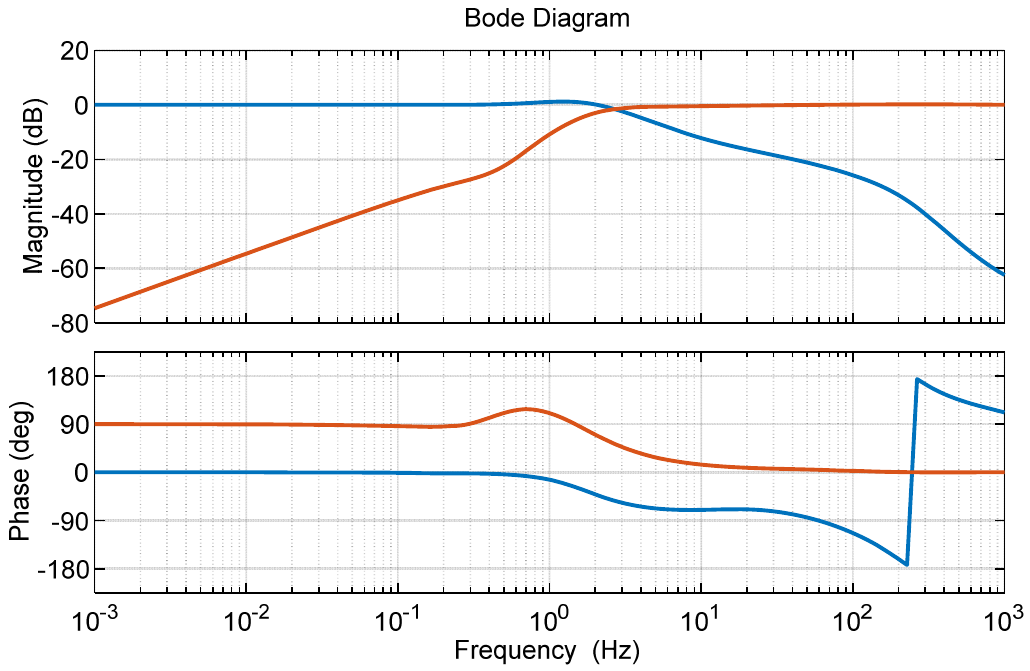}
	\vspace*{-2mm}
	\caption{Bode diagrams of $P_{53}(K)(s)$ and $P_{83}(K)(s)$ with droop controller in \eqref{eq:Frequency_droop_controller}. {\color{CBLUE}{\bf{-----}}} $P_{53}(K)(s)$ (i.e., complementary sensitivity function); {\color{CRED}{\bf{-----}}} $P_{83}(K)(s)$ (i.e., sensitivity function).}
	\vspace*{-2mm}
	\label{Fig_Power_Tracking}
\end{figure}

To begin with, we recall the concepts of {\em sensitivity function} and {\em complementary sensitivity function} \cite{skogestad2007multivariable}. The sensitivity function describes how the output rejects the disturbance from the input, while the complementary sensitivity function characterizes how the output tracks the input. Therefore, the Bode diagram of the sensitivity function is distinguished from that of the complementary sensitivity function. For example, Fig.~\ref{Fig_Power_Tracking} plots the Bode diagrams of $P_{53}(K)(s)$ and $P_{83}(K)(s)$ when the droop controller in \eqref{eq:Frequency_droop_controller} is applied. The transfer function $P_{53}(K)(s)$ describes how the active power tracks its reference $P_{\rm ref}$, i.e., a complementary sensitivity function. The transfer function $P_{83}(K)(s)$ describes how the active power error rejects the disturbance from $P_{\rm ref}$, i.e., a sensitivity function. It can be observed that the magnitude of the complementary sensitivity function approaches one ($0{\rm dB}$) in the low-frequency range and approaches zero in the high-frequency range. In contrast, the magnitude of the sensitivity function approaches one ($0{\rm dB}$) in the high-frequency range and approaches zero in the low-frequency range.

Based on the above characteristics of sensitivity function and complementary sensitivity function, we summarize the following two basic steps for choosing weighting functions to achieve reference tracking and disturbance rejection.

Step 1) Decide whether the output aims at tracking the input or rejecting the disturbance from the input. For instance, if the output aims at tracking the input, then the inverse of the weighting function should be the shape of a complementary sensitivity function, i.e., the magnitude approaches one in the low-frequency range and approaches zero in the high-frequency range.

Step 2) Decide the expected tracking/disturbance-rejection speed by assigning the bandwidth for the inverse of the weighting function.

For more details on choosing the weighting functions we refer to \cite{skogestad2007multivariable} and will further elaborate on the converter system below.

\subsection{Specification of the weighting functions for converters}
\label{weighting_function_PV}

\begin{figure}[!t]
	\centering
	\includegraphics[width=3.4in]{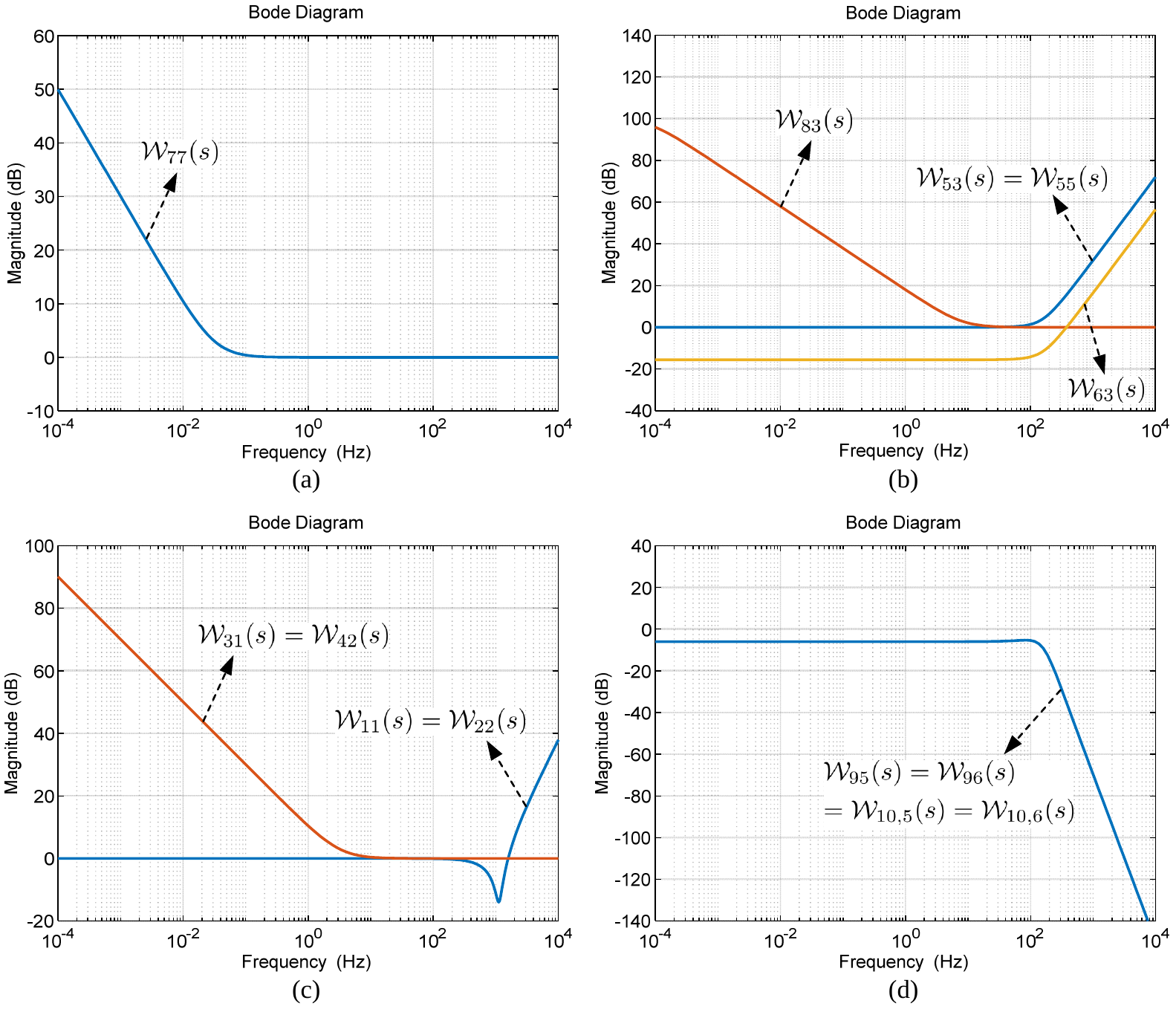}
	\vspace{-3mm}
	\caption{Magnitudes of the weighting functions (only the frequency range from $10^{-4}{\rm Hz}$ to $10^{4}{\rm Hz}$ is plotted).}
	\vspace{-4mm}
	\label{Fig_Weighting_Function}
\end{figure}

In the following, we list the control objectives of grid-connected power converters and discuss how these objectives can be achieved by the proper design of weighting functions, which follows from the two steps listed above in Section~\ref{subsec: weighting functions}.

i) Grid synchronization: As displayed in Fig.~\ref{Fig_Weighting_Function} (a), we choose $\mathcal W_{77}(s)$ as
\begin{equation}
\mathcal W_{77}(s) = \frac{s+s_{1{\_77}}}{s+s_{2{\_77}}}
\label{eq:W_AD}
\end{equation}
with $s_{1{\_77}}=0.2$ and $s_{2{\_77}}=0.0001$ such that the internal phase $\theta$ ($z_7$) has disturbance-rejection capability against $w_7$. According to Fig.~\ref{Fig_Angle_disturbance}, the converter can conveniently emulate grid-forming or grid-following (small-signal) dynamics by choosing different $\mathcal W_{77}(s)$ to shape $P_{77}(K)(s)$. Note that the choice of $\mathcal W_{77}(s)$ in Fig.~\ref{Fig_Weighting_Function} (a) will lead to grid-forming (small-signal) dynamics since the high gains appear only in the low-frequency range (hence the bandwidth of $P_{77}(K)(s)$ is limited).

ii) Power regulation: The active power should track its reference with fast dynamics, hence we choose $\mathcal W_{83}(s)$ as
\begin{equation}
\mathcal W_{83}(s) = \frac{s+s_{1{\_83}}}{s+s_{2{\_83}}}\,,
\label{eq:W_PD}
\end{equation}
where $s_{1{\_83}}=5$ and $s_{2{\_83}}=0.0005$ such that the tracking error is well eliminated in the low-frequency range, as displayed in Fig.~\ref{Fig_Weighting_Function} (b).
In addition, we choose $\mathcal W_{53}(s)$ as
\begin{equation}
\mathcal W_{53}(s) = \frac{s^2/\omega_{2{\_53}}^2 + 2\xi_{2{\_53}} s/\omega_{2{\_53}} + 1}{s^2/\omega_{1{\_53}}^2 + 2\xi_{1{\_53}} s/\omega_{1{\_53}} + 1}
\label{eq:W_PT}
\end{equation}
so as to suppress high-frequency disturbances for the power regulation. The magnitude of $\mathcal W_{53}(s)$ is displayed in Fig.~\ref{Fig_Weighting_Function} (b), with the parameters chosen as $\omega_{1{\_53}} = 7 \times 10^5$, $\xi_{1{\_53}} = 0.35$, $\omega_{2{\_53}} = 1000$ and $\xi_{2{\_53}} = 0.8$.
Moreover, we choose $\mathcal W_{55}(s)$ to be equal to $\mathcal W_{53}(s)$ such that the active power control is capable of rejecting disturbances from the power grid voltage.

Considering that when the terminal voltage remains constant, the change of active power flow also affects the steady-state value of the reactive power, we choose $\mathcal W_{63}(s)$ as
\begin{equation}
\mathcal W_{63}(s) = k_{\rm QT} \times \frac{s^2/\omega_{2{\_63}}^2 + 2\xi_{2{\_63}} s/\omega_{2{\_63}} + 1}{s^2/\omega_{1{\_63}}^2 + 2\xi_{1{\_63}} s/\omega_{1{\_63}} + 1}
\label{eq:W_QT}
\end{equation}
with $k_{\rm QT} = \frac{1}{6}$, $\omega_{1{\_63}} = 7 \times 10^5$, $\xi_{1{\_63}} = 0.35$, $\omega_{2{\_63}} = 1000$ and $\xi_{2{\_63}} = 0.8$. As shown in Fig.~\ref{Fig_Weighting_Function} (b), $\mathcal W_{63}(s)$ has a lower gain than $\mathcal W_{53}(s)$ in the low-frequency range due to the fact that the change of active power reference has less impact on the reactive power than on the active power, and the high gain of $\mathcal W_{63}(s)$ in the high-frequency range is used to suppress high-frequency disturbances.

iii) Voltage regulation: The {\em d}-axis and {\em q}-axis voltage components should track their reference values with fast dynamics, hence we choose $\mathcal W_{31}(s)$ and $\mathcal W_{42}(s)$ to be
\begin{equation}
\mathcal W_{31}(s) = \mathcal W_{42}(s) = \frac{s+s_{1{\_31}}}{s+s_{2{\_31}}}
\label{eq:W_VD}
\end{equation}
with $s_{1{\_31}} = 20$ and $s_{2{\_31}} = 10^{-5}$ in order to eliminate the tracking error in the low-frequency range, as shown in Fig.~\ref{Fig_Weighting_Function} (c). Furthermore, $\mathcal W_{11}(s)$ and $\mathcal W_{22}(s)$ are chosen to be
\begin{equation}
\mathcal W_{11}(s) = \mathcal W_{22}(s) = \frac{s^2/\omega_{2{\_11}}^2 + 2\xi_{2{\_11}} s/\omega_{2{\_11}} + 1}{s^2/\omega_{1{\_11}}^2 + 2\xi_{1{\_11}} s/\omega_{1{\_11}} + 1}
\label{eq:W_VT}
\end{equation}
with $\omega_{1{\_11}} = 7 \times 10^5$, $\xi_{1{\_11}} = 0.1$, $\omega_{2{\_11}} = 7 \times 10^3$ and $\xi_{2{\_11}} = 0.1$, as displayed in Fig.~\ref{Fig_Weighting_Function} (c). Note that $\mathcal W_{11}(s)$ and $\mathcal W_{22}(s)$ provide a magnitude drop around the resonance frequency of the {\em LCL} ($1100{\rm Hz}$ in this paper), which is used to reduce the control design emphasis around this frequency.

iv) Admittance performance: The transfer function matrix from $[w_5\;w_6]^\top$ to $[z_9\;z_{10}]^\top$ is $G(s)Y(s)$. This transfer function matrix will significantly affect the system stability as it incorporates the admittance matrix $Y(s)$. The choosing of $G(s)$ will be elaborated in the following section to show how it enforces decentralized stability. We choose the weighting functions for this transfer function matrix, i.e., $\mathcal W_{95}(s)$, $\mathcal W_{96}(s)$, $\mathcal W_{10,5}(s)$ and $\mathcal W_{10,6}(s)$, as
\begin{equation}
\begin{split}
&\;\mathcal W_{95}(s) = \mathcal W_{96}(s) = \mathcal W_{10,5}(s) = \mathcal W_{10,6}(s) \\
&= \frac{k_{\rm Ad}}{(s^2/\omega_{1{\_95}}^2 + 2\xi_{1{\_95}} s/\omega_{1{\_95}} + 1)^2}\,,
\end{split}
\label{eq:W_I}
\end{equation}
where $k_{\rm Ad} = 0.5$, $\omega_{1{\_95}}=1000$ and $\xi_{1{\_95}}=0.6$ in order to put more control effort in the frequency range of interest ($0\sim 100 {\rm Hz}$ in this paper). As displayed in Fig.~\ref{Fig_Weighting_Function} (d), $\mathcal W_{95}(s)$ has low gain in the high-frequency range to put less emphasis on shaping the high-frequency characteristics of $G(s)Y(s)$.

In addition to the above specified weighting functions in $\mathcal W(s)$, the other entries of $\mathcal W(s)$ are set as 0 such that the $\mathcal{H}_{\infty}$ problem focuses on the above design objectives.

\subsection{Constant reactive power control mode}
\label{weighting_function_PQ}

By employing the aforementioned setting, the converter regulates its active power and internal voltage magnitude (before the virtual impedance) to their reference values ($P_{\rm ref}$ and $V_d^{\rm ref}$), referred to as ``PV mode'' in this paper. In this mode, the reactive power is indirectly controlled by changing the reference of the internal voltage magnitude. As mentioned in Remark~\ref{Alternative_outputs}, the reactive power can also be directly controlled by choosing different integrals in $y$ such that the converter directly regulates the active and reactive power to their reference values, referred to as ``PQ mode''.

To be specific, the following steps are needed to change the system configuration and design $\mathcal{H}_{\infty}$-optimal controller for PQ mode: a) The output signal $y_2$ is changed to be the integral of reactive power tracking error (i.e., $Q_{\rm ref} - Q_E + w_4$) instead of the voltage magnitude tracking error ({\em d}-axis component). b) Let $\mathcal W_{11}(s) = \mathcal W_{31}(s) = 0$ be such that the $\mathcal{H}_{\infty}$ design does not impose voltage magnitude tracking. c) Let $z_3 = y_7$, $\mathcal W_{64}(s) = \mathcal W_{53}(s)$ and $\mathcal W_{34}(s) = \mathcal W_{83}(s)$ in order to achieve reactive power tracking and disturbance rejection, similar to the configuration for active power regulation as discussed in the previous subsection.

\section{Decentralized $\mathcal{H}_{\infty}$ Stability Certificates}
\label{Section Decentralized Stability Certificates}

In this section we present a decentralized $\mathcal{H}_{\infty}$ stability criterion and include it in the $\mathcal{H}_{\infty}$-control framework to ensure the stability of multi-converter systems through local design.

\subsection{Multi-device system descriptions}

Consider a Kron-reduced power network that interconnects $n$ devices, wherein the interior nodes are eliminated by assuming that the loads are constant current sources, similar to that in \cite{huang2019impacts}. Let $Y_i(s)$ denote the $2\times2$ admittance matrix of the $i{\rm th}$ device ($i \in \left\{1,...,n\right\}$). The terminal voltage and the current output of the $i{\rm th}$ device are respectively denoted by $U_i$ and $I_i$, which satisfies $I_i = Y_i(s)U_i$. Let ${\bf U},{\bf I} \in \mathbb{R}^{2n}$ be respectively the stacked voltage and current vectors of the $n$ devices, i.e., ${\bf U} = \left[U_1^\top\;...\;U_n^\top\right]^\top$ and ${\bf I} = \left[I_1^\top\;...\;I_n^\top\right]^\top$. The block diagonal admittance matrix ${\bf Y}(s) = {\rm diag}(Y_{1}(s),...,Y_{i}(s),...,Y_{n}(s))$ represents the dynamics of the $n$ devices, which satisfies
\begin{equation}
{\bf I} = {\bf Y}(s){\bf U}\,.
\label{eq:converter_dynamics}
\end{equation}

The transmission lines of the power network are assumed to be homogeneous with identical $R/L$ ratio. For a transmission line that connects node $i$ and node $j$ ($i,j \in \left\{{1,...,n} \right\},i \ne j$), the dynamic equation in the {\em dq} frame can be expressed as
\begin{equation}
\begin{split}
I_{ij} &= {B_{ij}}{{F}}(s)(U_i-U_j)\,,\\
{{F}}(s) &= \frac{{{1}}}{{{(s+\tau)^2/\omega _0} + \omega _0}}\left[ {\begin{array}{*{20}{c}}
	{s+\tau}&{ {\omega _0}}\\
	{ - {\omega _0}}&{s+\tau}
	\end{array}} \right]\,,		\label{eq:nodeij}
\end{split}
\end{equation}
where $I_{ij}$ is the current vector from node $i$ to node $j$, $U_i$ is the voltage at node $i$, $B_{ij} = 1/(L_{ij} \times \omega _0)$ is the susceptance between $i$ and $j$, { $\tau$ is the identical $R_{ij}/L_{ij}$ ratio of all the lines}, and $\omega_0$ denotes the nominal angular frequency \cite{huang2019impacts}.

Let $Q_{\rm red} \in \mathbb{R}^{n \times n}$ be the Kron-reduced grounded Laplacian matrix of the power network that encodes the line topology and susceptances \cite{dorfler2018electrical,dorfler2013kron}, calculated by
\begin{equation}
\begin{split}
{Q_{{\rm red}ij}} &=  - {B_{ij}},\;i \ne j\,,\\
{Q_{{\rm red}ii}} &= \sum\limits_{j = 1}^{n} {{B_{ij}}}\,,		\label{eq:Qmatrix}
\end{split}
\end{equation}
and $Q_{\rm red} \otimes {{F}}(s)$ is the corresponding admittance matrix of the network ($\otimes$ denotes the Kronecker product), which satisfies
\begin{equation}
{\bf I} = \left[Q_{\rm red} \otimes {{F}}(s)\right]{\bf U}\,,
\label{eq:network_admittance}
\end{equation}
or equivalently,
\begin{equation}
{\bf U} = \left[Q_{\rm red} \otimes {{F}}(s)\right]^{-1}{\bf I}\,.
\label{eq:network_impedance}
\end{equation}
Here $B_{ii}$ is the susceptance between node $i$ and the grounded node, i.e., {\em self-loop}. In this paper, the grounded node is the {\em infinite bus} (in small-signal modeling), similar to that in \cite{huang2019impacts}.

\begin{figure}[!t]
	\centering
	\includegraphics[width=3in]{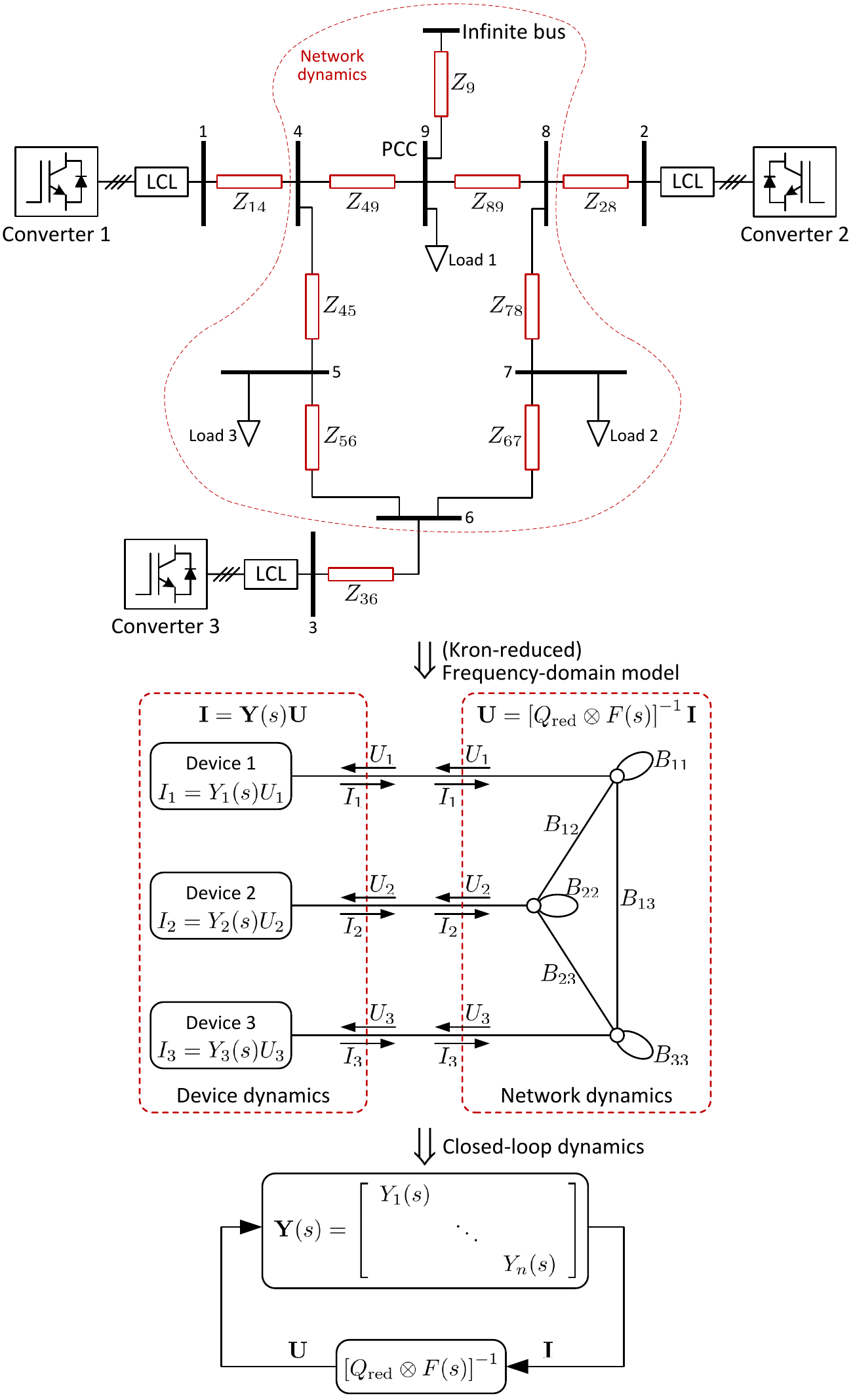}
	\vspace{-3mm}
	\caption{Closed-loop diagram of the multi-device system.}
	\vspace{-2mm}
	\label{Fig_multi_device_diagram}
\end{figure}

To better illustrate how the devices interact with the power network, Fig.~\ref{Fig_multi_device_diagram} shows a three-converter-nine-bus system that is connected to an infinite bus through the point of common coupling (the main parameters are given in Table~\ref{table:converter_parameter}). By using Kron reduction to eliminate the interior nodes and simplify the system modeling, an equivalent three-node network with self-loops can be obtained which interconnects the three converters (Nodes 4, 6 and 8 are remained; Nodes 5, 7, 9 and the infinite bus are eliminated; Nodes 1, 2 and 3 are included in the device dynamics) \cite{dorfler2013kron,huang2019impacts}. Note that the loads are modeled as constant current sources such that the interior nodes can be directly eliminated to simplify the analysis.
It can be seen that the input-output models of the devices and the network can be respectively represented by \eqref{eq:converter_dynamics} and \eqref{eq:network_impedance}, which together form the closed-loop system as depicted in Fig.~\ref{Fig_multi_device_diagram}. Therefore, the open-loop transfer function matrix of the multi-device system can be formulated by
\begin{equation}
{\bf L}(s) = -\left[Q_{\rm red} \otimes {{F}}(s)\right]^{-1}{\bf Y}(s)\,.
\label{eq:openloopeq}
\end{equation}

It can be seen that the closed-loop diagram of the multi-device system has a particular structure, that is, the device dynamics can be described by a block-diagonal matrix ${\bf Y}(s)$, and the admittance matrix can be formulated as the Kronecker product of $Q_{\rm red}$ and ${{F}}(s)$. In the following, we will show that the block diagonal structure of ${\bf Y}(s)$ enables decentralized stability certificates for multi-device systems.

\subsection{Decentralized $\mathcal{H}_{\infty}$ stability criterion}

We now present a decentralized $\mathcal{H}_{\infty}$ stability criterion for multi-device systems based on the small gain theorem.

\begin{proposition}[Stability of multi-device systems]
	\label{proposition: Heterogeneous dynamics}
	The multi-device system in \eqref{eq:openloopeq} is stable if
	\begin{equation}
	\mathop {\max}\limits_{i}\; \|F^{-1}(s)Y_{i}(s)\|_{\infty} < \lambda_1\,,
	\label{eq:small_gain}
	\end{equation}
	where $\|\cdot\|_{\infty}$ denotes the $\mathcal{H}_{\infty}$ norm, and $\lambda_1>0$ is the smallest eigenvalue of $Q_{\rm red}$.
	
	\begin{proof}
		According to the small gain theorem \cite{skogestad2007multivariable}, the system in \eqref{eq:openloopeq} is stable if
		\begin{equation}
		\|[Q_{\rm red} \otimes F(s)]^{-1}{\bf Y}(s)\|_{\infty} < 1\,.
		\label{eq:norm_openloop}
		\end{equation}
		
		Further, we have
		\begin{equation*}
		\begin{split}
		&\|[Q_{\rm red} \otimes F(s)]^{-1}{\bf Y}(s)\|_{\infty}\\
		=&\|[Q_{\rm red} \otimes {\mathcal I}_2]^{-1}[{\mathcal I}_n \otimes F^{-1}(s)]{\bf Y}(s)\|_{\infty}\\
		\le & \|[Q_{\rm red} \otimes {\mathcal I}_2]^{-1}\|_{\infty}  \times \|[{\mathcal I}_n \otimes F^{-1}(s)]{\bf Y}(s)\|_{\infty}\,,
		\end{split}
		\end{equation*}
		where ${\mathcal I}_n \in \mathbb{R}^{n \times n}$ denotes the $n$-dimensional identity matrix.
		
		Since
		\begin{equation*}
		\|[Q_{\rm red} \otimes {\mathcal I}_2]^{-1}\|_{\infty} = \bar \sigma(Q_{\rm red}^{-1}) = 1/\lambda_1
		\end{equation*}
		due to the symmetry of $Q_{\rm red}$, where $\bar \sigma(\cdot)$ denotes the largest singular value, and
		\begin{equation*}
		\begin{split}
		&\|[{\mathcal I}_n \otimes F^{-1}(s)]{\bf Y}(s)\|_{\infty} = \mathop {\max}\limits_{i}\; \|F^{-1}(s)Y_{i}(s)\|_{\infty}
		\end{split}
		\end{equation*}
		due to the block-diagonal structure of ${\bf Y}(s)$, it can then be deduced that condition \eqref{eq:norm_openloop} is satisfied if \eqref{eq:small_gain} holds, indicating that the system \eqref{eq:openloopeq} is stable, which concludes the proof.
	\end{proof}
\end{proposition}

Based on partitioning the system into two parts, which are the power network and the combination of all the devices as illustrated in Fig.~\ref{Fig_multi_device_diagram}, Proposition \ref{proposition: Heterogeneous dynamics} provides a convenient way to evaluate the overall system stability by simply looking at the network structure and the local dynamics of the devices. To be specific, the system is stable if $\|F^{-1}(s)Y_{i}(s)\|_{\infty}$ is smaller than $\lambda_1$ for every device (i.e., $ \forall i \in \left\{1,...,n\right\}$), while the system is more prone to instabilities if some devices present undesired dynamics characterized by high $\|F^{-1}(s)Y_{i}(s)\|_{\infty}$.

On the one hand, the transfer function matrix $F^{-1}(s)Y_{i}(s)$ in \eqref{eq:small_gain} is solely related to the local frequency-domain dynamics of the $i{\rm th}$ device, as $Y_i(s)$ is the device's admittance matrix and $F(s)$ is a fixed transfer function matrix determined by the line $R/L$ ratio. On the other hand, $\lambda_1$ is determined by the network structure, which in fact, can be seen as the connectivity strength of the network because $\lambda_1$ can only increase when the network becomes denser \cite{giordano2016smallest}. With Proposition \ref{proposition: Heterogeneous dynamics}, there are two ways to ensure the stability of the multi-device system. One is to make the power network dense enough (by building more transmission lines to make the grid strong enough) such that $\lambda_1$ is larger than $\|F^{-1}(s)Y_{i}(s)\|_{\infty}$ for every device, which however, is not practical and needs lots of investments.

The other way is to decrease $\|F^{-1}(s)Y_{i}(s)\|_{\infty}$ of the devices such that it is smaller than $\lambda_1$. Note that $\lambda_1$ is a constant if the network structure remains unchanged. Once $\|F^{-1}(s)Y_{i}(s)\|_{\infty} < \lambda_1$ is satisfied for every device, the overall system is stable, that is, by Proposition \ref{proposition: Heterogeneous dynamics}, the stability of the multi-device system can be guaranteed in a decentralized way.
We remark that by choosing $G(s)=F^{-1}(s)$ in \eqref{eq:z}, $\|F^{-1}(s)Y(s)\|_{\infty}$ can be conveniently decreased via the $\mathcal{H}_{\infty}$-control design proposed in Section \ref{section Hinf controller}, as $F^{-1}(s)Y(s)$ is a submatrix of $P(K)(s)$.

\subsection{$\mathcal{H}_{\infty}$-design with decentralized stability certificates}

Recall that Proposition \ref{proposition: Heterogeneous dynamics} provides a sufficient condition to evaluate the system stability by partitioning the multi-device system into the network part and the device part. In order to reduce the conservativeness of this condition, we include the grid impedance $L_g$ ($L_g$ consists of grid-side inductor of the {\em LCL} and line inductance) in the device part when partitioning the system, which increases $\lambda_1$ of the network part.


\renewcommand\arraystretch{1.25}
\begin{table}
	\scriptsize
	\centering
	\caption{Parameters of the Power Converter and Power Network}
	\begin{tabular}{|ll|}
		\hline
		\multicolumn{2}{|c|}{Base values for per-unit calculation}										\\
		\hline
		Voltage base value: $U_{\rm{b}} = 380~\rm{V}$	&Power base value:	$S_{\rm{b}} = 50~\rm{kVA}$	\\
		Frequency base value: $f_{\rm{b}} = 50~\rm{Hz}$	&												\\
		\hline
		\multicolumn{2}{|c|}{Parameters of the Power Part (per-unit values)}							\\
		\hline
		Converter-side inductor: $L_F = 0.05$			&\textit{LCL} capacitor:	$C_F = 0.05$		\\
		Grid-side inductor: $L_g = 0.05$				&$R/L$ ratio of the grid impedance: $\tau = 0.1$\\
		\hline
		\multicolumn{2}{|c|}{Parameters of the Control Part}											\\
		\hline
		\multicolumn{2}{|l|}{PI gains of the current control loop: $0.5~{\rm{p.u.}},10~{\rm{p.u.}}$}	\\
		\multicolumn{2}{|l|}{Voltage feedforward control: $K_{\rm VF} = 1,T_{\rm VF} = 0.004~{\rm s}$}			\\
		\multicolumn{2}{|l|}{Virtual reactance: $X_v = 0.3~{\rm{p.u.}}$}			\\
		\hline
		\multicolumn{2}{|c|}{Network Parameters (per-unit values)}										\\
		\hline
		\multicolumn{2}{|c|}{$Z_{14} = 0.015 + j0.15$, $Z_{49} = 0.00125 + j0.0125$, $Z_{89} = 0.0025 + j0.025$}\\
		\multicolumn{2}{|c|}{$Z_{28} = 0.015 + j0.15$, $Z_{78} = 0.012 + j0.12$, $Z_{67} = 0.008 + j0.08$}\\
		\multicolumn{2}{|c|}{$Z_{45} = 0.0012 + j0.012$, $Z_{56} = 0.0007 + j0.007$, $Z_{36} = 0.015 + j0.15$}\\
		\multicolumn{2}{|c|}{$Z_{9} = 0.0005 + j0.005$, $P_{\rm Load1} = 0.12, P_{\rm Load2} = 0.2, P_{\rm Load3} = 0.18$}\\
		\hline
	\end{tabular}		
	\vspace{-1mm}
	\label{table:converter_parameter}
\end{table}

In this manner, the converter's grid impedance does not have to be known exactly as it contains the line inductance. Therefore, we consider the following $\mathcal{H}_{\infty}$-optimal control problem to ensure the robustness against various grid inductances
\begin{equation}
	\begin{split}
		&\mathop {\min }\limits_K \mathop {\max }\limits_{L_g \in \mathbb{L}} \|{\mathcal W}(s) \circ P(K)(s)\|_{\infty}\\
		=& \mathop {\min }\limits_K \mathop {\max }\limits_{L_g \in \mathbb{L}} \mathop {\max }\limits_\omega \bar \sigma\left[{\mathcal W}(j\omega) \circ P(K)(j\omega)\right]\,,
		\label{eq:robust_Xg}
	\end{split}
\end{equation}
where the set $\mathbb{L} = \left\{L|L=0.05x,x \in \{1,...,10\}\right\}$.

\begin{figure}[!t]
	\centering
	\includegraphics[width=2.5in]{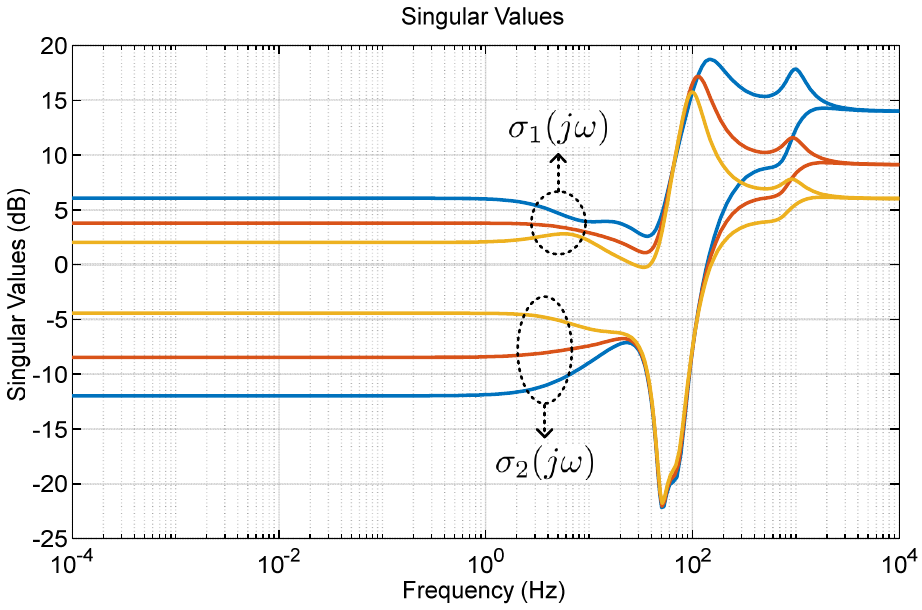}
	\vspace{-2mm}
	\caption{Singular values of $F^{-1}(j\omega)Y(j\omega)$ with $\mathcal{H}_{\infty}$-optimal controller in \eqref{eq:optimal_K}: {\color{CBLUE}{\bf{-----}}} $L_g=0.2~{\rm p.u.}$, {\color{CRED}{\bf{-----}}} $L_g=0.35~{\rm p.u.}$, {\color{CYELLOW}{\bf{-----}}} $L_g=0.5~{\rm p.u.}$.}
	\label{Fig_SV_FY_Hinf}
\end{figure}

Given the weighting functions in Section \ref{weighting_function_PV} (PV mode) and the system parameters in Table \ref{table:converter_parameter}, we use the {\em hinfstruct} routine in \text{MATLAB} to solve the $\mathcal{H}_{\infty}$-optimal control problem in \eqref{eq:robust_Xg}, and the static gain matrix $K$ is obtained as
\begin{equation}
	\begin{split}
		&K = \\
		&{\tiny
			\left[ {\begin{array}{*{20}{c}}
					{-0.01}&{392.7}&{0.1}&{183.3}&{0}&{38.8}&{-2.1}\\
					{-3.01}&{-67.1}&{-0.09}&{484.6}&{0}&{-62.9}&{4.8}\\
					{-97.7}&{-1.9}&{-134.5}&{2.3}&{55.7}&{-0.4}&{0.04}
			\end{array}} \right]}.
	\end{split}
	\label{eq:optimal_K}
\end{equation}

The corresponding singular values of the transfer function matrix $F^{-1}(j\omega)Y(j\omega)$ (denoted by $\sigma_1(j\omega)$ and $\sigma_2(j\omega)$, $\sigma_1(j\omega) \ge \sigma_2(j\omega)$) are plotted in Fig.~\ref{Fig_SV_FY_Hinf}. It can be seen that there is no unexpected resonance peak, and for the three choices of the grid inductance, it holds that $\sigma_1(j\omega)<20{\rm dB}=10, \forall \omega$. Moreover, $\sigma_1(j\omega)$ has a lower magnitude in the frequency range below $100{\rm Hz}$ due to the choice of the weighting function in \eqref{eq:W_I}, which enhances the robustness of the system and prevents sub-synchronous oscillations \cite{liu2017subsynchronous}.

Additionally, an $\mathcal{H}_{\infty}$-optimal controller for PQ mode can be obtained by carrying out the three steps in Section \ref{weighting_function_PQ} and then solving \eqref{eq:robust_Xg}, which leads to the  static gain matrix
\begin{equation}
\begin{split}
&K = \\
&{\tiny
	\left[ {\begin{array}{*{20}{c}}
		{-1.35}&{-61.8}&{0.66}&{361.8}&{0}&{13.5}&{0}\\
		{-0.77}&{-46.1}&{-0.22}&{-27.2}&{0}&{-14.9}&{-0.02}\\
		{-0.3}&{-9.3}&{-257.5}&{-8.3}&{61.6}&{-2.5}&{0.95}
		\end{array}} \right]}.
\end{split}
\label{eq:optimal_K_PQ}
\end{equation}

\section{Simulation Results}
\subsection{Single-converter system}

\begin{figure}[!t]
	\centering
	\includegraphics[width=3.2in]{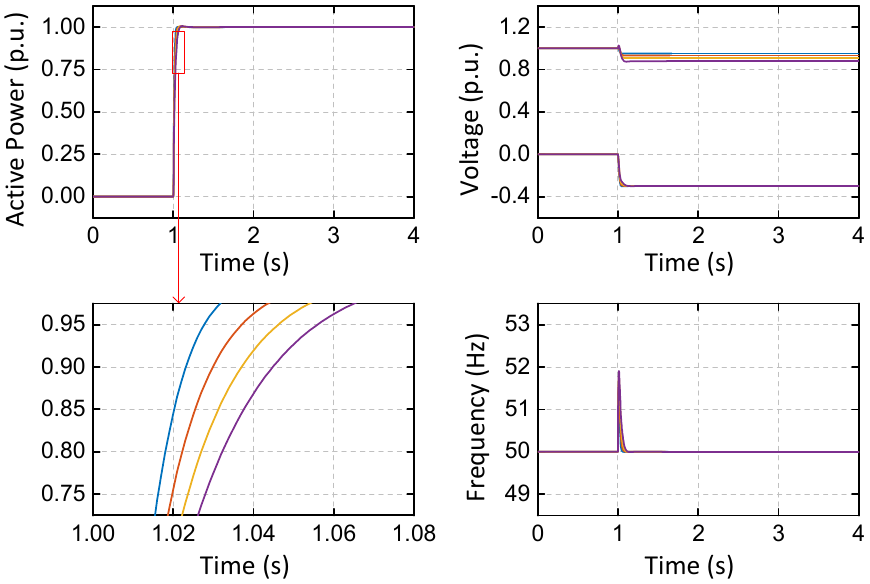}
	\vspace{-2mm}
	\caption{Time-domain responses of the single-converter system with $\mathcal{H}_{\infty}$-optimal controller: {\color{CBLUE}{\bf{-----}}} $L_g=0.05~{\rm p.u.}$, {\color{CRED}{\bf{-----}}} $L_g=0.2~{\rm p.u.}$, {\color{CYELLOW}{\bf{-----}}} $L_g=0.35~{\rm p.u.}$, {\color{CPURPLE}{\bf{-----}}} $L_g=0.5~{\rm p.u.}$}
	\vspace{0mm}
	\label{Fig_responses_Hinf}
\end{figure}

To illustrate the effectiveness of the $\mathcal{H}_{\infty}$-optimal design, we now provide detailed simulation studies based on the nonlinear model of the single-converter system in Fig.~\ref{Fig_Control_diagram}. The converter parameters are given in Table \ref{table:converter_parameter}, and the $\mathcal{H}_{\infty}$-optimal controller has been obtained in the previous section.

Fig.~\ref{Fig_responses_Hinf} shows the time-domain responses with the $\mathcal{H}_{\infty}$-optimal controller applied. At $t=1~{\rm s}$, the active power reference steps from $0$ to $1.0~{\rm p.u.}$. It can be seen that even with different grid inductances $L_g$, the active power has nearly the same response (fast dynamics and no overshoot). We note that the {\em d}-axis and {\em q}-axis voltage components have different steady-state values due to the virtual inductance, because the voltage controller regulates the virtual voltage behind the virtual inductor to the reference values. The internal frequency of the converter also has fast responses and the anticipated performance obtained through the weighting functions.

\begin{figure}[!t]
	\centering
	\includegraphics[width=3.4in]{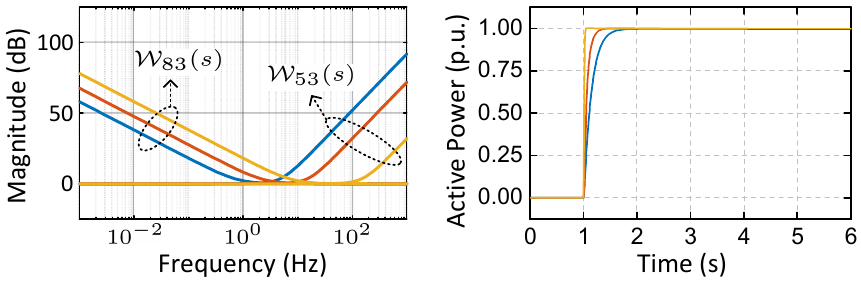}
	\vspace{-2mm}
	\caption{Time-domain responses of the active power with different weighting function designs in \eqref{eq:W_PD} and \eqref{eq:W_PT}: {\color{CBLUE}{\bf{-----}}} $s_{1\_83} = 1.2,~\omega_{2\_53} = 12.9$, {\color{CRED}{\bf{-----}}} $s_{1\_83} = 5,~\omega_{2\_53} = 31.6$, {\color{CYELLOW}{\bf{-----}}} $s_{1\_83} = 25,~\omega_{2\_53} = 316.2$.}
	\label{Fig_responses_Hinf_P}
\end{figure}

Instead of directly changing control parameters to achieve different dynamic performances, the $\mathcal{H}_{\infty}$-design specifies different weighting functions to achieve different performances. For example, if we want to change the tracking speed of the active power, we need to correspondingly change the shapes of $W_{83}(s)$ and $W_{53}(s)$ in \eqref{eq:W_PD} and \eqref{eq:W_PT}. Fig.~\ref{Fig_responses_Hinf_P} plots $W_{83}(s)$ and $W_{53}(s)$ with different bandwidths (the blue ones have the lowest bandwidths and the yellow ones have the highest bandwidths). By solving the $\mathcal{H}_{\infty}$-optimal control problem in \eqref{eq:robust_Xg} one obtains different $\mathcal{H}_{\infty}$-optimal controller $K$. Fig.~\ref{Fig_responses_Hinf_P} shows the time-domain responses of the active power when the different weighting functions are adopted, and it can be seen that all the responses have the anticipated performance (no overshoot). Moreover, increasing the bandwidths of $W_{83}(s)$ and $W_{53}(s)$ leads to a faster response of the active power, that is, the $\mathcal{H}_{\infty}$ design provides a convenient and systematic way to achieve expected system dynamics.

\begin{figure}[!t]
	\centering
	\includegraphics[width=2.75in]{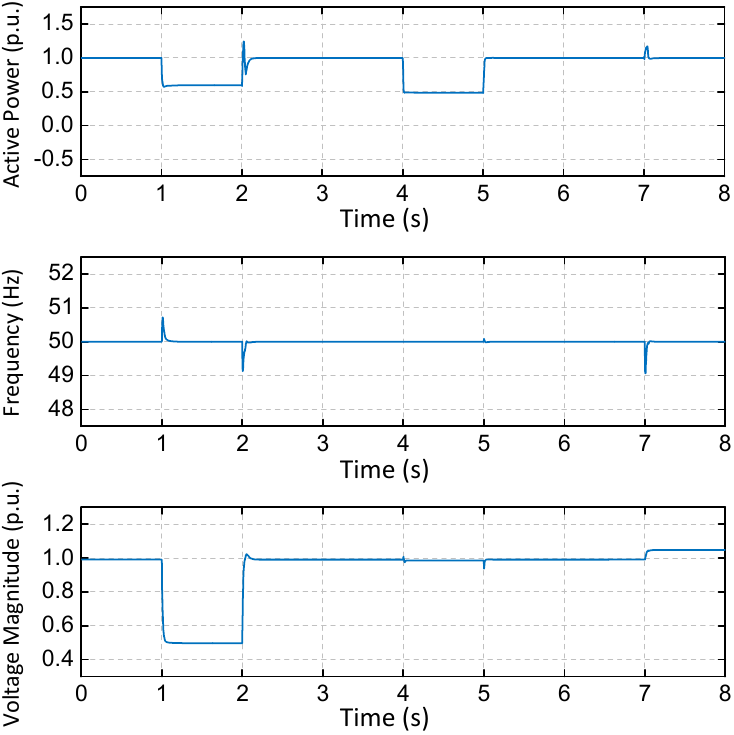}
	\vspace{-2mm}
	\caption{Responses of the $\mathcal{H}_{\infty}$-optimal controller under disturbances (the current reference $I^{\rm ref}_{{\rm C}d}$ is limited within $\pm 1.1~{\rm p.u.}$ by a saturation link, and $I^{\rm ref}_{{\rm C}q}$ is limited within $\pm 0.5~{\rm p.u.}$ such that voltage support can be provided under disturbances and the current magnitude is limited within about 1.2~p.u.).}
	\label{Fig_responses_Hinf_disturbances}
\end{figure}

We also test the transient (large-signal) performance of the $\mathcal{H}_{\infty}$-optimal controller, as shown in Fig.~\ref{Fig_responses_Hinf_disturbances}. The grid voltage magnitude drops from 1~p.u. to 0.5~p.u. at $t = 1{\rm s}$ and recovers to 1~p.u. at $t = 2{\rm s}$. The converter is disconnected from the grid at $t = 4{\rm s}$ and reconnected to the grid at $t = 5{\rm s}$ (the converter is stilled attached to a local load ($P_{\rm Load} = 0.5~{\rm p.u.}$) when disconnected from the grid). The internal voltage reference $V_d^{\rm ref}$ (before the virtual inductance) steps from 1~p.u. to 1.45~p.u. at $t = 7{\rm s}$, which makes the converter's voltage magnitude (at the {\em LCL}'s capacitor point) change to about 1.05~p.u. because the virtual inductance takes up most of the voltage drop from the internal voltage to the grid voltage.
	
Overall, the converter has acceptable transient performance under the aforementioned severe disturbances, even without additional control switching or auxiliary loops. Note that sometimes auxiliary loops are needed to help the converter ride through faults and large disturbances, e.g., in conventional droop-controlled converters \cite{zhang2010modeling,huang2017transient}. 
It is in fact  also possible to fix the structure of $K$ (by forcing some elements to be zeros) before solving \eqref{eq:robust_Xg} in order to make the resulting controller share the same structure with some widely-used controllers, e.g., droop controller, such that the experience on transient behavior analysis and auxiliary loop design can be inherited.

\begin{figure}[!t]
	\centering
	\includegraphics[width=3.2in]{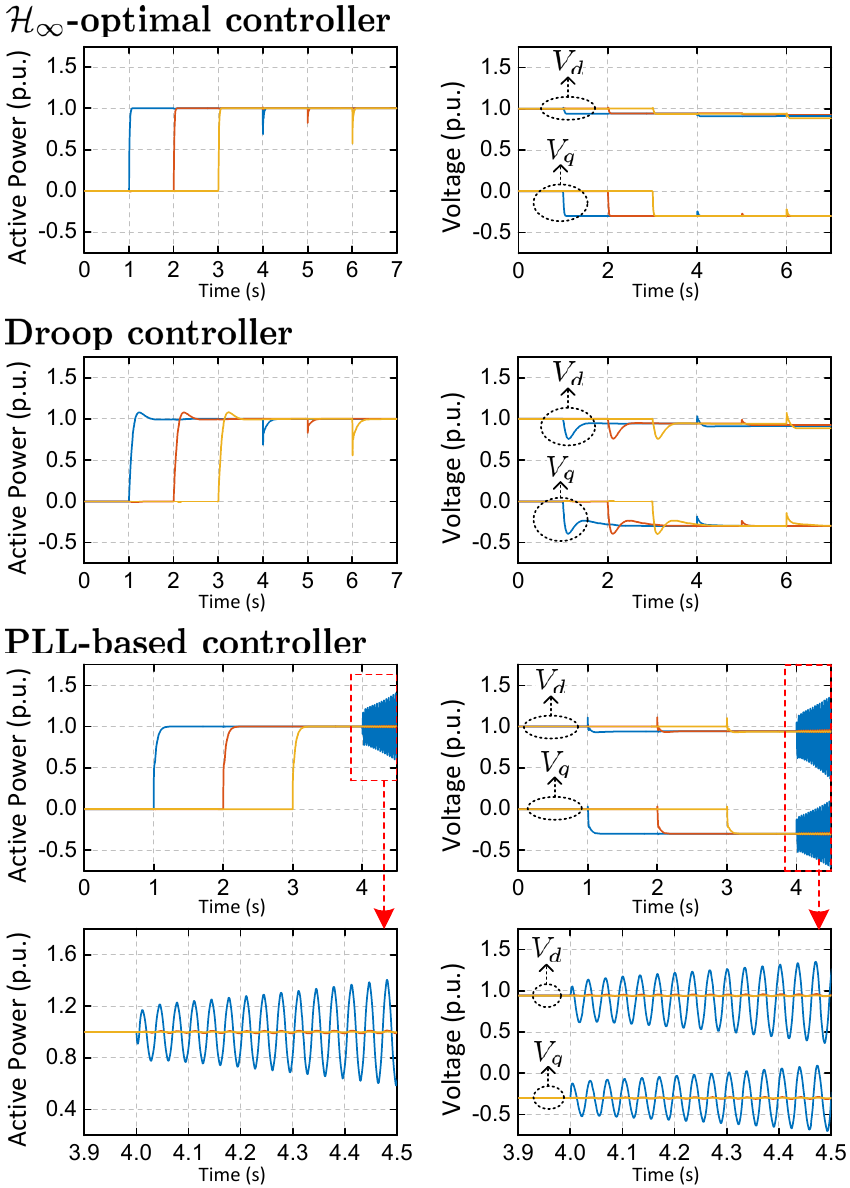}
	\vspace{-2mm}
	\caption{Responses of the three-converter system: {\color{CBLUE}{\bf{---}}} Converter 1, {\color{CRED}{\bf{---}}} Converter 2, {\color{CYELLOW}{\bf{---}}} Converter 3. The active power references of the three converters step from $0$ to $1.0~{\rm p.u.}$ at $t=1~{\rm s}$, $t=2~{\rm s}$ and $t=3~{\rm s}$, respectively. To simulate the effects of changes of grid topology, e.g., line outages, the inductance $L_{g1}$ steps from $0.2~{\rm p.u.}$ to $0.4~{\rm p.u.}$ at $t=4~{\rm s}$, $L_{g2}$ steps from $0.2~{\rm p.u.}$ to $0.3~{\rm p.u.}$ at $t=5~{\rm s}$, and $L_{g3}$ steps from $0.2~{\rm p.u.}$ to $0.5~{\rm p.u.}$ at $t=6~{\rm s}$.}
	\vspace{-3mm}
	\label{Fig_responses_multiConverter}
\end{figure}

\subsection{Three-converter system}
\label{subsec: three-converter system}

In what follows, we test the performance of the $\mathcal{H}_{\infty}$-optimal controller in the three-converter system in Fig.~\ref{Fig_multi_device_diagram}. The grid inductances of the converters are $L_{g1} = L_{g2} = L_{g3} = 0.2~{\rm p.u.}$ (e.g., $L_{g1}$ includes the grid-side inductance of the {\em LCL} filter of Converter~1 and the inductance in $Z_{14}$).
The Kron-reduced Laplacian matrix of the network is
\begin{equation*}
Q_{\rm red} =
\scriptsize
\left[ {\begin{array}{*{20}{c}}
	{114.55}&{-10}&{-54.55}\\
	{-10}&{40}&{-5}\\
	{-54.55}&{-5}&{59.55}
	\end{array}} \right]\,,
\end{equation*}
whose smallest eigenvalue is $\lambda_1 = 21.11$. It can be deduced that the condition in \eqref{eq:small_gain} is satisfied because $\|F^{-1}(s)Y(s)\|_{\infty}<\lambda_1$ as shown in Fig.~\ref{Fig_SV_FY_Hinf}, indicating that the stability of the three-converter system is guaranteed when applying the $\mathcal{H}_{\infty}$-optimal controller. Although the stability condition in \eqref{eq:small_gain} is a sufficient one, its conservativeness in practice is acceptable. For example, with $L_{g1} = L_{g2} = L_{g3} = 0.5~{\rm p.u.}$, the three-converter system is stable if $\lambda_1 > 15.8{\rm dB}=6.17$ (deduced from Fig.~\ref{Fig_SV_FY_Hinf}). By changing the network parameters, we observed from the simulation results that the real stability boundary is about $\lambda_1 = 5$, which is remarkably close to $6.17$. Moreover, the condition for $\lambda_1$ (with regards to the results in Fig.~\ref{Fig_SV_FY_Hinf}) is reasonable because the long transmission line that interconnects the device (e.g., some remote renewable base) and the grid can be included in the device side and further robustified in \eqref{eq:robust_Xg}. In summary, our sufficient and decentralized small-gain condition \eqref{eq:small_gain} is remarkably tight for the considered case study (which we attribute to optimal design incorporating the condition \eqref{eq:small_gain}). Moreover, the minor conservativeness serves as a robustness margin to model uncertainties considering the fact that the system may still be stable even if the condition is not satisfied.

Fig.~\ref{Fig_responses_multiConverter} plots the time-domain responses of the three-converter system with the $\mathcal{H}_{\infty}$-optimal controller, which shows that the system presents the anticipated performance (fast dynamics and no overshoot) to disturbances such as changes of power reference and grid topology.
For comparison, Fig.~\ref{Fig_responses_multiConverter} also shows the responses when the three-converter system applies the droop controller in \eqref{eq:Frequency_droop_controller} and the PLL-based controller in \eqref{eq:PLL_based_controller}. It can be seen that the droop controller has overshoots and presents slower dynamics under the changes of grid topology. The PLL-based controller has no overshoot in the responses to the power reference steps, but the system becomes unstable when $L_{g1}$ steps from $0.2~{\rm p.u.}$ to $0.4~{\rm p.u.}$ at $t=4~{\rm s}$ (i.e., when the grid becomes weaker), consistent with the prevailing intuition that conventional grid-following converters are stable only in strong grids.
The above results demonstrate the superiority of the $\mathcal{H}_{\infty}$-optimal controller over the conventional droop controller in \eqref{eq:Frequency_droop_controller} and the PLL-based controller in \eqref{eq:PLL_based_controller}.

\subsection{HIL real-time simulation results}

\begin{figure}[!t]
	\centering
	\includegraphics[width=3in]{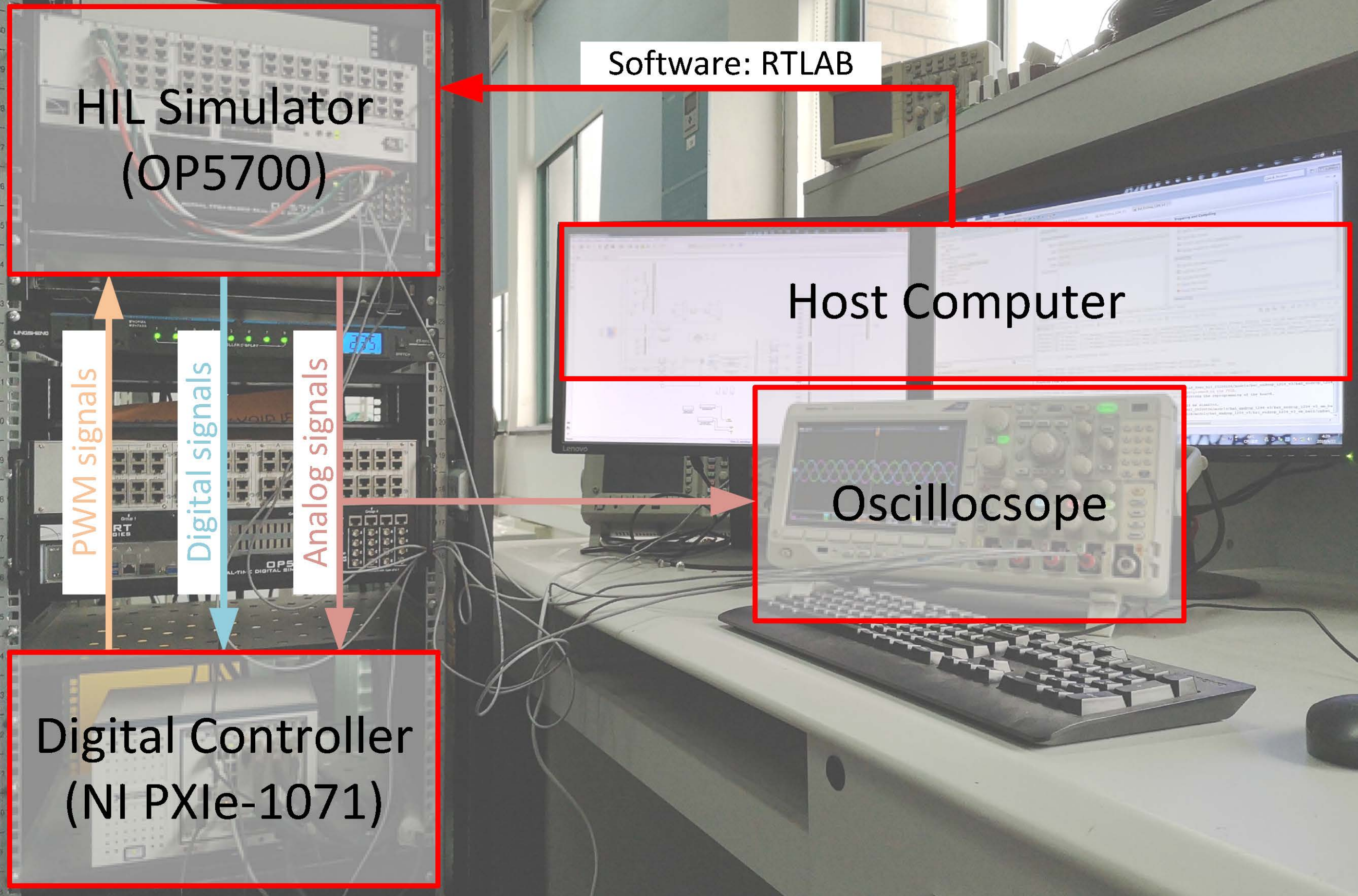}
	\vspace{-2mm}
	\caption{HIL real-time simulation platform.}
	\label{Fig_HIL_platform}
\end{figure}

We now illustrate the effectiveness of the $\mathcal{H}_{\infty}$-optimal controller using detailed and high-fidelity HIL real-time simulations. The HIL platform is shown in Fig.~\ref{Fig_HIL_platform}, which comprises an HIL simulator (OP5700), a digital controller (NI PXIe-1071), a host computer, and an oscilloscope. The HIL simulator has CPU and FPGA resources for real-time calculations, and is equipped with inputs/outputs to communicate with the digital controller. Here we use the FPGA resource to simulate the power part of the three-converter system in Fig.~\ref{Fig_multi_device_diagram} in real time, which allows small step size (less than 0.5~$\mu$s) and thus high accuracy. The $\mathcal{H}_{\infty}$-optimal controller of Converter~1 is implemented in the digital controller (NI PXIe-1071) to test its performance in practice, and we independently separate two CPU cores in the HIL simulator to realistically implement the $\mathcal{H}_{\infty}$-optimal controllers for Converter~2 and Converter~3. Here the three converters are operated in PQ mode by employing the controller obtained in \eqref{eq:optimal_K_PQ}.

\begin{figure}[!t]
	\centering
	\includegraphics[width=2.8in]{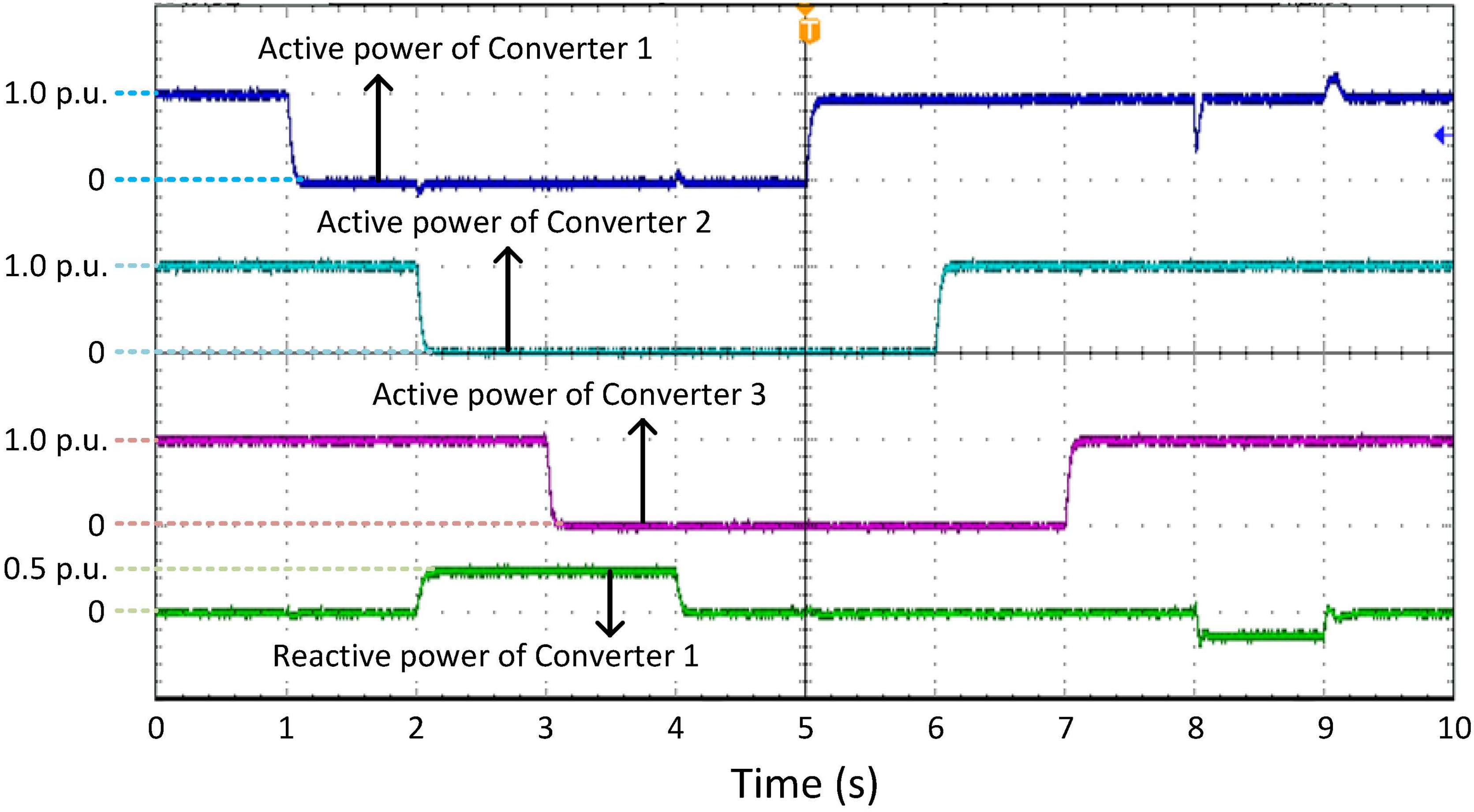}
	\vspace{-2mm}
	\caption{Time-domain responses of the three-converter system. The active power reference of Converter~1 (2, 3) steps from 1~p.u. to 0 at $t = 1~{\rm s}$ ($t = 2~{\rm s}$, $t = 3~{\rm s}$) and back to 1~p.u. at $t = 5~{\rm s}$ ($t = 6~{\rm s}$, $t = 7~{\rm s}$). The reactive power reference of Converter~1 steps from 0 to 0.5~p.u. at $t = 2~{\rm s}$ and back to 0 at $t = 4~{\rm s}$. Converter~1 is switched from PQ mode to PV mode at $t = 8~{\rm s}$ and back to PQ mode at $t = 9~{\rm s}$.}
	\label{Fig_HIL_PQ}
\end{figure}

\begin{figure}[!t]
	\centering
	\includegraphics[width=3.5in]{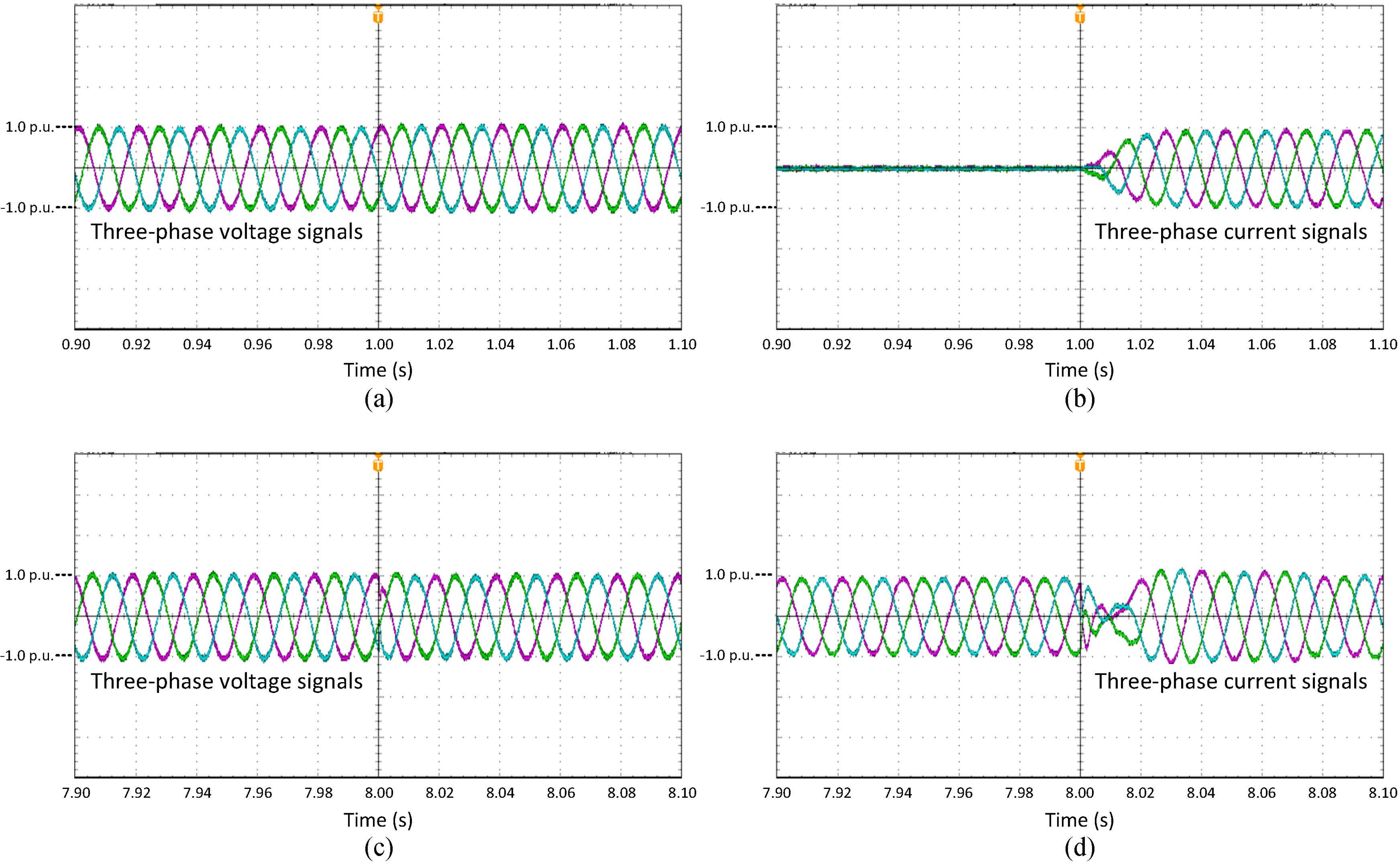}
	\vspace{-6mm}
	\caption{Three-phase voltage and current waveform of Converter~1. (a) Voltage waveform at around 1~s. (b) Current waveform at around 1~s. (c) Voltage waveform at around 8~s. (d) Current waveform at around 8~s.}
	\label{Fig_HIL_VI}
\end{figure}

Fig.~\ref{Fig_HIL_PQ} and Fig.~\ref{Fig_HIL_VI} display the time-domain responses of the three-converter system obtained from the HIL real-time simulations. It can be seen from Fig.~\ref{Fig_HIL_PQ} that the converters track the active and reactive power references with fast dynamics (the rising time is less than 0.1~s) and no overshoot, which again confirms and validates the effectiveness of our earlier control design and theoretical analysis. The transient performance under control mode switching (between PQ mode and PV mode) is also fast and smooth. Fig.~\ref{Fig_HIL_VI} shows the three-phase voltage and current waveform of Converter~1 at around 1~s (active power reference step) and at around 8~s (control mode switching), which demonstrates that the three-phase voltages are very well maintained under the disturbances, and the currents have fast dynamics and acceptable transient performance.

\section{Conclusions}

This paper proposed an $\mathcal{H}_{\infty}$-control design framework for grid-connected power converters to perform robust and optimal control. Instead of tuning parameters based on eigenvalue analysis or engineering experience, the proposed $\mathcal{H}_{\infty}$-control is a systematic way to achieve optimal performance in terms of the multiple control objectives of power converters. We illustrated how the converter can be specified as grid-forming or grid-following type with regards to small-signal dynamics by properly choosing the weighting functions. Moreover, we first proposed a decentralized stability criterion for multi-device systems which demonstrates how to ensure the global stability of the entire system by local control design. We further presented how this decentralized stability certificate can be included in the $\mathcal{H}_{\infty}$-control design to guarantee the stability of multi-converter systems. The obtained $\mathcal{H}_{\infty}$-optimal controller was tested by detailed simulations and HIL implementation of a three-converter system, which showed that the $\mathcal{H}_{\infty}$-optimal controller presents the anticipated dynamic performance and robust stability against variable grid conditions.

\bibliographystyle{IEEEtran}

\bibliography{ref}

\end{document}